\documentclass[lettersize,journal]{IEEEtran}

\usepackage{amsmath,amsfonts,amsthm}
\usepackage{algorithmic}
\usepackage{graphicx}
\usepackage{textcomp}
\usepackage{xcolor}
\usepackage{pifont}
\usepackage{enumitem}
\usepackage{balance}
\usepackage{booktabs}
\usepackage{arydshln}
\usepackage{threeparttable}
\usepackage{makecell}
\usepackage{colortbl}
\usepackage{xspace}
\usepackage[many]{tcolorbox}
\usepackage{hyperref}
\usepackage{tikz}
\usepackage{gensymb}
\usepackage{multirow}
\usepackage{subcaption}
\usetikzlibrary{decorations.shapes}

\tikzset{
	decoration = {shape backgrounds, shape=circle,
		shape size=1pt, shape sep={4pt, between centers}
	},
	paint/.style = {decorate, fill=black}
}

\newtcolorbox{findingbox}{%
    left=3pt,
    right=3pt,
    top=0pt,
    bottom=0pt,
    boxrule=0.3mm,
    colframe=black!75!white,
    colback=white,
    breakable
}%

\newtcolorbox{implicationbox}{%
    left=3pt,
    right=3pt,
    top=0pt,
    bottom=0pt,
    boxrule=0mm,
    colback=black!10!white,
    breakable,
    frame empty
}%

\newcolumntype{L}[1]{>{\raggedright\let\newline\\\arraybackslash\hspace{0pt}}m{#1}}
\newcolumntype{C}[1]{>{\centering\let\newline\\\arraybackslash\hspace{0pt}}m{#1}}
\newcolumntype{R}[1]{>{\raggedleft\let\newline\\\arraybackslash\hspace{0pt}}m{#1}}

\theoremstyle{definition}
\newtheorem{finding}{Finding}
\newtheorem{implication}{Implication}

\newcounter{subfidcounter}

\newcounter{subimpcounter}

\newcommand{\totalReviewAmount}{1,656,968\xspace}

\newcommand{\totalAppAmount}{28,745\xspace}

\newcommand{\updatedReviewSampleAmount}{6,925\xspace}
\newcommand{\temporalSamplePerQuarter}{2,000\xspace}

\newcommand{\reviewSampleSize}{7,000\xspace}
\newcommand{\informativeReviewAmount}{1,554\xspace}

\newcommand{\manualAnalysisIterationNum}{four\xspace}

\newcommand{\qualityFirstLayerCategoryAmount}{12\xspace}

\newcommand{\factorFirstLayerCategoryAmount}{ten\xspace}

\makeatletter
\def\adl@drawiv#1#2#3{%
	\hskip.5\tabcolsep
	\xleaders#3{#2.5\@tempdimb #1{1}#2.5\@tempdimb}%
	#2\z@ plus1fil minus1fil\relax
	\hskip.5\tabcolsep}
\newcommand{\cdashlinelr}[1]{%
	\noalign{\vskip\aboverulesep
		\global\let\@dashdrawstore\adl@draw
		\global\let\adl@draw\adl@drawiv}
	\cdashline{#1}
	\noalign{\global\let\adl@draw\@dashdrawstore
		\vskip\belowrulesep}}
\makeatother

\newcommand{\mytodocomment}[2]{}

\definecolor{mygray}{gray}{.9}
\definecolor{applegreen}{rgb}{0.55, 0.71, 0.0}
\definecolor{alizarin}{rgb}{0.82, 0.1, 0.26}
\definecolor{cadmiumorange}{rgb}{0.93, 0.53, 0.18}
\definecolor{carrotorange}{rgb}{0.93, 0.57, 0.13}

\let\diff\undefined %

\ifdefined\diff
\newcommand{\textred}[1]{\textcolor{red}{#1}}
\newcommand{\textblue}[1]{\textcolor{blue}{#1}}

\else
\newcommand{\textred}[1]{#1}
\newcommand{\textblue}[1]{#1}

\fi

\makeatletter
\def\namedlabel#1#2{\begingroup
	#2%
	\def\@currentlabel{#2}%
	\phantomsection\label{#1}\endgroup
}
\makeatother

\def\BibTeX{{\rm B\kern-.05em{\sc i\kern-.025em b}\kern-.08em
    T\kern-.1667em\lower.7ex\hbox{E}\kern-.125emX}}
\begin{document}

\title{\textred{
Modeling Software Quality in Virtual Reality Applications from User Feedback}}

\author{Shuqing~Li,        
	Chenran~Zhang,
	Yepang~Liu,
	Cuiyun~Gao,
	Shing-Chi~Cheung,
	and~Michael~R.~Lyu
\IEEEcompsocitemizethanks{
\IEEEcompsocthanksitem Shuqing~Li, and Michael~R.~Lyu are with The Chinese University of Hong Kong, Hong Kong, China. E-mail:\{sqli21, lyu\}@cse.cuhk.edu.hk
\IEEEcompsocthanksitem Yepang~Liu is affiliated with both the Research Institute of Trustworthy Autonomous Systems and the Department of Computer Science and Engineering at Southern University of Science and Technology, Shenzhen, China. E-mail: liuyp1@sustech.edu.cn
\IEEEcompsocthanksitem Chenran~Zhang and Cuiyun~Gao are with Harbin Institute of Technology, Shenzhen, China. E-mails: 220110514@stu.hit.edu.cn and gaocuiyun@hit.edu.cn
\IEEEcompsocthanksitem Shing-Chi~Cheung is with Hong Kong University of Science and Technology, Hong Kong, China. E-mail:scc@cse.ust.hk}
}

\markboth{\textred{IEEE Transactions on Software Engineering}}%
{\textred{Li \MakeLowercase{\textit{et al.}}: Modeling Software Quality in Virtual Reality Applications from User Feedback}}
\IEEEtitleabstractindextext{%

\begin{abstract}
    \textred{Virtual Reality (VR) applications couple software behavior with head-mounted displays, body movement, spatial interaction, and multimodal feedback. Consequently, familiar software-quality problems can have distinct consequences in VR, yet empirical knowledge of the quality concerns expressed by VR users remains fragmented.}

		\textred{We analyze \totalReviewAmount public reviews of \totalAppAmount VR applications across seven app stores. Our semiautomatic workflow uses five unsupervised methods to generate candidate aspects and multi-round human open and focused coding to construct and refine a hierarchical model of user-perceived VR software quality.}
    \textred{The resulting model comprises \qualityFirstLayerCategoryAmount quality attributes and \factorFirstLayerCategoryAmount influencing factors. It distinguishes conventional attributes whose consequences change in VR from attributes closely tied to VR, including multisensory perception, user-friendly interaction mechanisms, immersivity, and comfort and safety. The influencing factors connect perceived quality problems to movement, control, tracking, multimedia, physics, configuration, and content decisions.}
		\textred{Supplementary analyses examine temporal variation, platform and genre composition, accessibility-related reviews, and developer update logs. These descriptive analyses show that the concerns surfaced by reviews depend on context and identify concrete priorities for VR quality assurance.}
		\textred{We release the available datasets, taxonomy materials, and analysis results to support replication and follow-up research: \href{https://sites.google.com/view/vr-software-quality}{\color{blue}{https://sites.google.com/view/vr-software-quality}}.}
\end{abstract}

\begin{IEEEkeywords}
	Virtual Reality Applications, Software Quality, User Reviews, Empirical Study.
\end{IEEEkeywords}
}

\maketitle

\IEEEdisplaynontitleabstractindextext

\IEEEpeerreviewmaketitle

\section{Introduction}
\label{sec:introduction}

\IEEEPARstart{V}{irtual} Reality (VR) offers immersive experiences by creating interactive virtual environments for users.
It has been used for various purposes, such as virtual collaboration~\cite{website:xr-application-AltspaceVR}, military training~\cite{paper:vr-military-training}, medical operations~\cite{paper:vr-ar-in-surgery}, and entertainment~\cite{website:xr-application-xrgames}.
More than 15,000 
VR applications have been deployed to various app stores, including Oculus App Store~\cite{website:oculus-app-store}, Steam~\cite{website:steam-app-store-vr}, and VIVEPORT~\cite{website:viveport}, which have attracted over 171 million users worldwide~\cite{website:vr-users-number}. 
The Metaverse vision has driven a surge in the VR application market. 
The worldwide market size of VR achieved a substantial \$28.41 billion in 2022 and is projected to undergo rapid expansion at a compound annual growth rate of 13.8\%~\cite{website:vr-market-gvr}.
Given that VR industry is still in its infancy and keeps growing rapidly, understanding users' expectations for the software quality of VR applications is crucial for effective quality assurance practices and market success.
VR applications differ from traditional software in many aspects.
For example, they rely on the revolutionized software-user interaction mechanisms to provide users with immersive experiences.
Users need to use special devices and their body movements to control the movement and actions of their avatars and interact with other objects in the immersive three-dimensional virtual worlds.
This poses greater challenges to VR development 
and makes users have higher and more diverse expectations on the software quality~\cite{book:sqa} of VR applications, as compared with traditional software.
In particular, some issues that are tolerable in traditional software may become severe in VR applications.
Let's take performance inefficiency~\cite{standard:iso-25010-square-2011} as an example.
While using traditional software with poor performance, the time delay often just causes users' impatience and affects their user experience at most, sometimes it is even unnoticeable when the time delay is not so large.
However, for highly immersive VR experiences, low-performance situations with stuttering graphics or even a relatively low-level time delay can break the senses of immersion, cause significant discomfort to users, and potentially result in motion sickness.
Here is a user complaint on the severe motion sickness caused by a low-performance popular VR game, \textit{Arizona Sunshine}~\cite{website:vr-game-arizona-Sunshine}, ``\textit{The dropping framerate and the lag from poor RAM performance ... give me the worst motion sickness I have ever gotten in VR\footnote{Modified and simplified to ease understanding. Raw reviews are on our website.}}''.
\textblue{Prior VR software studies have examined focused concerns such as ad fraud~\cite{paper:webvr-ad-attack-lee21}, privacy-policy compliance~\cite{paper:ovrseen-Trimananda22}, performance optimization~\cite{paper:vr-performance-optimization}, and game-design complaints~\cite{paper:vrgame-empirical-Epp21}. Other reviews synthesize expert-defined metrics and evaluation methods rather than quality concerns expressed in deployed applications. What remains missing is an empirically grounded model that connects the quality attributes raised in public VR-app feedback to the design and implementation factors that users associate with those attributes across multiple app-store ecosystems.}
This raises the barrier to effective VR quality assurance and improvement.
\textblue{To address this gap, we empirically model user-perceived software quality in VR applications. Here, the \emph{user perspective} is deliberately operationalized through public app-store reviews; it complements, but does not represent, developer, expert, telemetry, or organizational perspectives.}
We focus on the following two important research questions: 

\textbf{\textit{What software quality attributes are of major concern to VR application users?
What are the major factors that influence user experience on the most concerned quality attributes?}}
\textblue{Public VR-app reviews offer direct, large-scale evidence of experiences, expectations, and requirements that reviewers choose to report~\cite{paper:app-love-users-ase15, paper:recommend-software-changes-fse16}. They do not establish the identity, representativeness, or motivation of every reviewer, a limitation we address in Section~\ref{sec:threats}.}
\textblue{For our study, we collect \totalReviewAmount public reviews of \totalAppAmount
VR applications from seven VR-related app stores.}
\textblue{To answer our research questions, we mine fine-grained concepts about software quality
from the large corpus of reviews to construct a taxonomy of quality attributes and analyze the corresponding influencing factors.}
\textblue{The approach is semiautomatic by design: NLP methods generate a broad set of candidate aspects at corpus scale, whereas human qualitative coding makes the conceptual decisions required to define, relate, and refine taxonomy categories.}
\textblue{Via review analysis, we construct a taxonomy of \qualityFirstLayerCategoryAmount software quality attributes expressed in public VR-app feedback (Section~\ref{sec:answers-to-rqs}).
The frequently represented VR-specific attributes concern multisensory perception, interaction mechanisms, immersivity, and comfort and safety. These attributes make explicit how VR's embodied and immersive interfaces change the consequences of software-quality decisions.
}
\textblue{The taxonomy supplies a review-grounded vocabulary for defining VR quality requirements and organizing quality-assurance activities.}
\textblue{We further identify \factorFirstLayerCategoryAmount influencing factors that connect quality concerns to concrete design and implementation decisions, including movement and control mechanisms, real-life motion tracking, multimedia systems, physics, configuration, and content. These connections turn the attribute taxonomy from a list of concerns into an evidence base for prioritizing development, testing, and future research (Section~\ref{sec:answers-to-rqs}).}
\textblue{To summarize, our work makes the following contributions:
\begin{itemize}[leftmargin=*, topsep=2pt, itemsep=2pt]
		\item A cross-ecosystem empirical characterization of user-perceived VR software quality, based on \totalReviewAmount public reviews of \totalAppAmount VR applications from seven app stores.
		\item A hierarchical model comprising \qualityFirstLayerCategoryAmount quality attributes and \factorFirstLayerCategoryAmount influencing factors, together with an explicit account of how VR's embodied and immersive properties extend or alter conventional software-quality concerns.
    We connect these attributes to actionable development and quality-assurance targets and triangulate them through descriptive temporal, ecosystem, accessibility-related, genre, and update-log analyses.
		\item Public datasets, taxonomy materials, and analysis results that support inspection, replication, and follow-up research at \href{https://sites.google.com/view/vr-software-quality}{\color{blue}{https://sites.google.com/view/vr-software-quality}}.
\end{itemize}
}

\section{Background}

\subsection{Virtual Reality Applications}
\label{subsec:background-vr-software}
Most VR applications are developed using 3D content creation frameworks, such as Unity~\cite{website:unity} and Unreal Engine~\cite{website:unreal-engine}.
Although different frameworks may employ different implementations, or even different languages for development (e.g., C\# for Unity and C++ for Unreal Engine), the terms and concepts for the construction of VR applications are similar~\cite{paper:vr-performance-optimization, website:unreal-engine-doc-for-unity-dev}.
In the rest of this paper, we will use the terms and concepts in Unity for illustration.

Despite the potentially divergent underlying framework, a VR application is usually composed of multiple interconnected \textit{scenes} for users to immerse in.
Users can navigate to and explore different \textit{scenes}.
There are structured virtual \textit{objects} in each \textit{scene}, many of which are interactable.
Specific runtime behaviors are defined in source code files, usually known as \textit{scripts}.
The predefined \textit{lifecycle methods} will be executed by framework during each frame to perform 
\textit{physics updates} (ensuring accurate and consistent simulation of physical interactions, such as the collision of two objects or the application of gravitational force on an entity), user input handling, \textit{logic updates} (managing game states and other non-physics related changes, like updating a player's score or changing the state of a game level based on certain conditions), and \textit{rendering}.
Besides \textit{scripts}, there are also various types of \textit{asset files} in VR projects, including \textit{3D model files}, \textit{texture files}, \textit{audio files}, \textit{text files}, \textit{plug-ins and code-related assets} and others, to create and enrich application content.

\subsection{VR Devices \& Interaction}
\label{subsec:vr-interaction}
Typical VR applications can be accessed through (1) PC-based VR (PCVR) device sets (which require computer connections for computation resources), (2) standalone VR device sets (which have built-in processors and can work without computers), or (3) mobile phone based VR device sets (which use mobile phones for computation and display).
Different from traditional software, VR applications support user interactions through more devices apart from common ones such as keyboards, mice, and cameras.
They often have full support for various special devices designed for creating immersive experiences.
We call such special devices ``\textit{VR devices}'' in this paper for ease of presentation.

There are three typical types of VR devices, including:
\begin{itemize}[leftmargin=*, topsep=2pt, itemsep=2pt]
	\item (a) \textit{Head-mounted displays} (HMD, a.k.a. headsets), which display digital image frames and play sound;
	\item (b) \textit{Handheld controllers} (a.k.a. gamepads), which can be used to control VR applications through spatial movement and button pressing, usually one for each hand;
	\item (c) \textit{Position tracking equipment} (sensors, cameras, or other locating devices), which works with HMD and tracks the body movement of users, equipping VR applications with more immersive exploration capacities of six degrees of freedom (DoF).
\end{itemize}

Besides, new devices are emerging:
\begin{itemize}[leftmargin=*, topsep=2pt, itemsep=2pt]
    \item (d) \textit{Advanced wearable VR devices}, such as gloves~\cite{website:metagloves}, wearable haptic suits~\cite{website:bhaptics, website:skinetic, website:teslasuit}, cyber shoes~\cite{website:cybershoes}, etc.
    \item (e) \textit{Specially shaped VR devices dedicated for several genres of applications}, including steering wheel controllers~\cite{website:thrustmaster-steering-wheel-pedal}, pedals~\cite{website:thrustmaster-steering-wheel-pedal}, gun stocks~\cite{website:protubevr-gun-stock}, etc.
\end{itemize}
The advanced and dedicated devices usually help users to interact with VR applications in an easier and more comfortable way using their body movement, or convey more dimensions of feedback from applications to users.
Such feedbacks are typically haptic senses provided through vibrations of vibro-tactile motors~\cite{website:bhaptics, website:skinetic}, electrical stimulation of electro-tactile electrodes~\cite{website:teslasuit} or the shapes of devices (dedicated devices).

\section{Related Work}
\subsection{Studies on VR Applications}

Most existing work on VR applications focuses on understanding or mitigating some specific issues, 
including prevention from ad fraud attacks~\cite{paper:webvr-ad-attack-lee21}, privacy policy auditing~\cite{paper:ovrseen-Trimananda22}, current practices of performance optimization~\cite{paper:vr-performance-optimization}, game design~\cite{paper:vrgame-empirical-Epp21}\textblue{, and stereoscopic visual inconsistencies that can induce cybersickness~\cite{Li2024StereoVR}}.

Multiple techniques were proposed to ensure security and privacy of VR applications.
Luo et al.~\cite{paper:oculock-luo20} proposed to 
use the human visual system to build an unobservable VR authentication system.
Lee et al.~\cite{paper:webvr-ad-attack-lee21} analyzed possible third-party ad fraud attacks on web-based VR applications, and further proposed AdCube to prevent such threats with small latency.
Trimananda et al.~\cite{paper:ovrseen-Trimananda22} proposed OVRseen for network traffic and privacy policy auditing in Oculus VR.
\textblue{More recently, Guo et al.~\cite{Guo2025MetaVRSecurity} combined static, taint, privacy-policy, and user-review analyses to study security and privacy risks in 900 Meta VR applications.}

Several empirical studies have been conducted to explore and understand VR applications.
Rodriguez and Wang~\cite{paper:os-vr-rodriguez17} studied the growing trends, popular topics, and common file structures of open-source VR projects.
Adams et al.~\cite{paper:sec-interview-adams18} conducted interviews with VR developers and users to understand their concerns on security and privacy.
Li et al.~\cite{paper:issre-webxr-empirical} performed an empirical study on web-based extended reality bugs to understand their symptoms, root causes, and uniqueness.
Nusrat et al.~\cite{paper:vr-performance-optimization} analyzed the optimization commits in open-source Unity-based VR projects to better understand VR performance optimization.
Rzig et al.~\cite{paper:vr-testing-empirical} conducted a quantitative and qualitative empirical study to analyze existing testing practices of open-source Unity-Based VR applications.
\textblue{Li et al.~\cite{Li2026MultiUserXRDefects} subsequently characterized defects in multi-user XR systems, including their symptoms, root causes, and repair strategies.}
\textblue{These studies focused on a specific topic, while we aim to systematically analyze quality concerns of VR applications across multiple dimensions.}

\textblue{An early review-based study by Epp et al.~\cite{paper:vrgame-empirical-Epp21} analyzed Steam VR games and player complaints for game-design insight, emphasizing broad categories such as lacking content, game-specific issues, and community concerns. Recent work has expanded review-based XR analysis in several directions. Dong et al.~\cite{Dong2024MultilingualVRReviews} mine multilingual Steam reviews to identify what VR-game players praise or criticize. Zaman et al.~\cite{Zaman2025MobileXRReviews} extract feature concerns and suggestions from mobile XR reviews. Guo et al.~\cite{Guo2025MetaVRSecurity} use reviews as one source of evidence about security and privacy concerns, alongside program and policy analyses. Li et al.~\cite{Li2025XRCybersickness} use large-scale XR app-store reviews to assess cybersickness and reason about its causes.}

\textblue{These studies establish review-based XR analysis as an active research line rather than an unstudied area. Their empirical targets remain task- or context-specific, such as one game marketplace, mobile feature requests, security and privacy, or cybersickness. Our study addresses a different analytical object: a cross-marketplace, multidimensional model that jointly organizes user-perceived VR software-quality attributes and the design or implementation factors associated with them. This relationship-centered structure provides an organizing context for more specialized findings.}

 \subsection{App Review Analysis}

 App stores provide a channel for users to share their feedback as well as learn others' experiences.
 These reviews are valuable for both users (including current and prospective ones) and developers.
 Although app reviews offer an opportunity for developers to improve their projects, it requires massive efforts of them to analyze these explosive amount of reviews manually.
\textblue{
Because app-review research is extensive, we discuss representative studies most directly related to extracting quality concepts and development-relevant information rather than attempting an exhaustive survey. This literature includes work on feedback classification, change-request extraction, rationale mining, review-based user stories, emerging-issue detection, and review selection~\cite{paper:app-review-classification-ase15, paper:review-classification-re15, paper:recommend-software-changes-fse16, paper:change-requests-icse17, paper:mining-user-rationale-re17, paper:review-user-stories-guo20, paper:emerging-issue-detection-tse21, paper:which-user-review-TSE21}. Recent work augments review classification with issue-report data~\cite{Abedini2024AppReviewClassification} and uses LLMs to translate competitor reviews into feature-improvement suggestions~\cite{Assi2026LLMCure}. These studies establish app reviews as useful software-engineering evidence, but their units of analysis are usually review-mining tasks or conventional mobile and game software rather than VR software-quality models. Our study instead mines public app-store feedback from VR applications across seven stores to construct joint hierarchical taxonomies of software-quality attributes and their influencing factors, and relates these taxonomies to cross-ecosystem evidence and supplementary analyses.
}

 Many prior works focus on proposing a review analysis approach for a specific task~\cite{paper:app-love-users-ase15, paper:change-requests-icse17, paper:idea-online-app-review-icse18}.
 Chen et al.~\cite{paper:ar-miner-icse14} proposed AR-Miner to help developers discover informative user reviews.
 Gu and Kim~\cite{paper:app-love-users-ase15} presented a review summarization and visualization framework, named SUR-Miner.
 Palomba et al.~\cite{paper:change-requests-icse17} proposed CHANGEADVISOR, in order to help mobile app developers find and localize user change requests.
 Recently, Tushev et al.~\cite{paper:keyatm-icse22} utilized keyword-assisted topic models to extract useful information from mobile app reviews.
 Various studies were also made to leverage app review analysis methods, either automatically or manually, to extract useful information for further analysis.
 OASIS was proposed by Wei et al.~\cite{paper:oasis-fse17} to prioritize static analysis warnings for Android applications based on user reviews.
 Both Gui et al.~\cite{paper:ad-review-analysis-gui17} and Gao et al.~\cite{paper:ad-review-analysis-gao22} conducted empirical studies on mobile app reviews that are related to in-app ads to study ad-related app issues.
 Lin et al.~\cite{paper:steam-game-reviews-empirical-lin19} performed review analysis on Steam games to understand user concerns about games,
 thus helping developers improve their games. %

\textblue{These methods inform our candidate-discovery stage, but our research question additionally requires qualitative decisions about concept boundaries, hierarchy, and relationships between attributes and influencing factors. We therefore use NLP methods to generate candidate aspects at corpus scale and human open and focused coding to construct and refine the final model.}
\textblue{\subsection{Metrics and Evaluation Methods for VR Applications}}

\textblue{
Prior work also studies how VR quality and experience can be measured. Samini and Palmerius summarize performance metrics for VR/AR interaction evaluation~\cite{8120326}; Kong and Liu discuss strategies and metrics for VR quality of experience~\cite{10.1007/978-3-030-02053-8_48}; Ankomah and Vangorp classify VR research by metrics such as presence, immersion, interaction, performance, and usability~\cite{eb18e2448bef4b66a216cf376a0be521}; and Kuri et al. map software-quality metrics used for VR products~\cite{Kuri2021VRQualityMetrics}. Ramaseri-Chandra and Reza survey 188 respondents about preselected VR quality attributes and identify usability and performance as the most highly ranked~\cite{RamaseriChandra2024VRQualityAttributes}. Questionnaire- and human-factor-oriented reviews further organize subjective instruments and factor measurements for VR experiences~\cite{BAREISYTE2024100505, info16010035}.}

\textblue{
Another line of secondary studies reviews evaluation and testing methods. Domain-focused reviews summarize how immersive VR is evaluated in higher education and social VR~\cite{RADIANTI2020103778, 10.1145/3711078}; Peres et al. review methods for evaluating immersive 3D virtual environments~\cite{Peres2024Immersive3DEvaluationMethods}; and Gu et al. map XR software-testing research, including testing activities, concerns, techniques, and evaluation methodologies~\cite{Gu2025XRTestingMapping}. Recent primary studies operationalize particular test targets: VRExplorer models VR scenes for semi-automated exploration~\cite{Zhu2025VRExplorer}, while Li et al. improve context-sensitive GUI grounding for automated XR testing~\cite{Li2026XRGUI}.}

\textblue{
Together, this literature shows growing interest in both VR/XR quality assurance and app-store feedback. The empirical objects are complementary: metric and review papers synthesize instruments or testing facets, while recent review-based studies target a particular application type, quality concern, user group, or decision task. Our work constructs a cross-marketplace, multidimensional model from public reviews and jointly organizes quality attributes and their influencing factors, thereby connecting user-perceived outcomes to actionable design and implementation concerns.
}

\section{Methodology}
\label{sec:methodology}

 \begin{figure}[t!] 
 	\centering 
 	\includegraphics[width=\columnwidth]{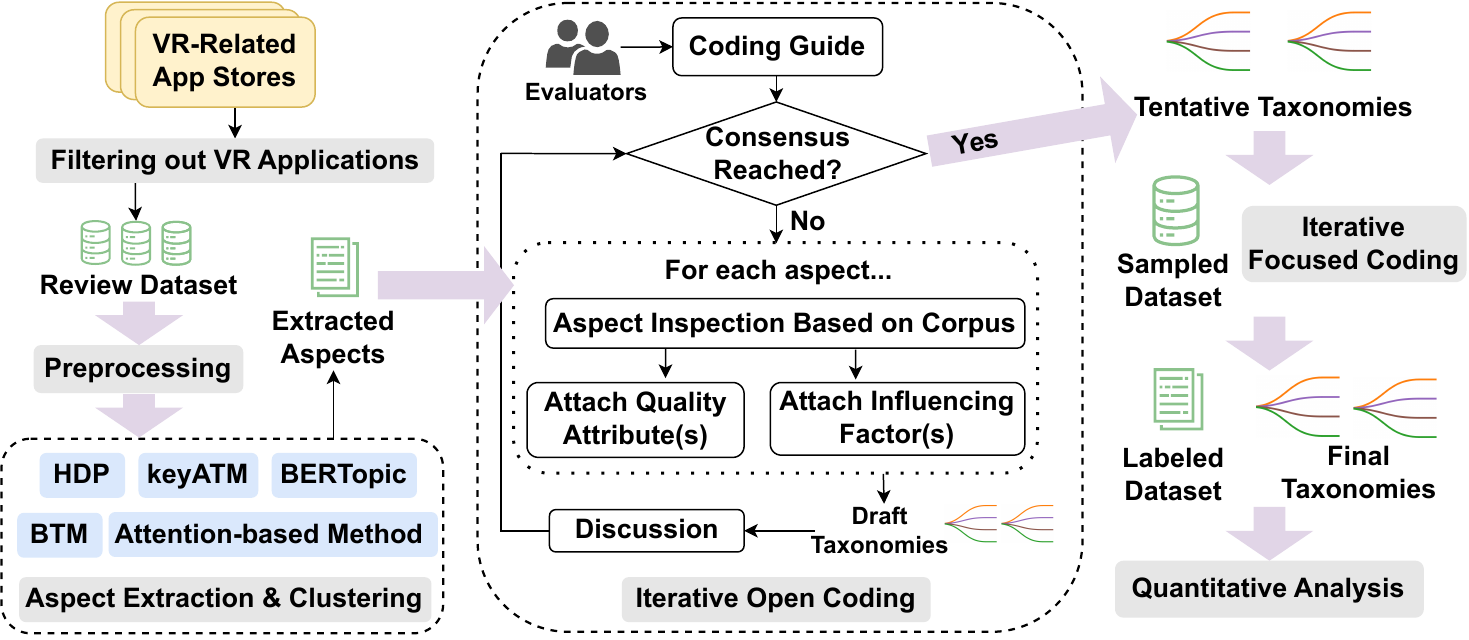} 
 	\caption{Overview of Methodology}
 	\label{fig:overview}
 \end{figure}

\textblue{We investigate the software-quality attributes expressed in public VR-application reviews and the design or implementation factors that reviewers associate with those attributes. The complete category sets are outcomes of the analysis described below, not a predefined selection of examples. These outputs answer the two research questions introduced in Section~\ref{sec:introduction}.}
\textblue{In this paper, a \textit{user} denotes a person or account that posted a public review for a collected VR application. The \textit{user perspective} denotes quality concerns observable from these public review texts; it does not include developer telemetry, expert inspection, private support tickets, or direct demographic verification unless otherwise stated. Because app stores do not expose verifiable reviewer identities or motivations, we cannot guarantee that every review is genuine or non-manipulated. We therefore treat reviews as public app-store feedback and discuss potential review manipulation as a threat to validity rather than as something eliminated by preprocessing or supplementary labeling.}
\textblue{To better answer these research questions, we build a semiautomatic approach that combines automatic NLP candidate generation with human conceptual coding. The NLP stage scales the search for candidate aspect terms across a large corpus and reduces dependence on a single extraction model, while the manual coding stage determines whether candidates are meaningful quality attributes, influencing factors, or irrelevant/noisy terms. Figure~\ref{fig:overview} shows the overview. We first collect a large-scale VR review dataset from selected app stores (Section~\ref{subsec:method-dataset-collection}). Upon preprocessing (Section~\ref{subsubsec:methodology-preprocessing}), we use multiple NLP approaches to extract commonly mentioned candidate aspect terms from review texts automatically.}
\textblue{Based on the extracted candidate aspects, we then conduct iterative open coding~\cite{book:open-coding-16} for constructing the taxonomies of quality attributes and influencing factors (Section~\ref{subsec:taxonomy-construction}).}
\textblue{After that, we follow the focused coding procedure~\cite{book:sage-qda-focused-coding} to further validate and refine our taxonomies on a sampled dataset, and conduct quantitative analysis on the resulting labeled dataset}
\textblue{(Section~\ref{subsec:method-manual-analysis}).}

\subsection{Dataset Collection}
\label{subsec:method-dataset-collection}

\textblue{We collected reviews from seven publicly accessible marketplaces: four VR-specific stores, Steam, Google Play, and the iOS App Store. We selected these sources to cover standalone, PC-assisted, and mobile distribution while retaining stores with sufficient public feedback for analysis (at least ten applications with more than ten reviews). Table~\ref{table:xrzoo} reports the resulting corpus.}
\textblue{For VR-specific stores, we collected all applications and English reviews available through the store interfaces at collection time.}
\textblue{Because Steam and the mobile stores also distribute non-VR applications, we used marketplace-specific inclusion rules. For Steam,}
\textblue{we included applications officially tagged ``\textit{VR Only}.'' The mobile stores do not expose an equivalent VR-support tag, so we used the following procedures:}
\textblue{(1) For Google Play, we first searched using \textit{VR} and \textit{virtual reality}, then}
\textblue{used LibScout~\cite{paper:libscout} to identify applications containing one of the targeted Android VR SDKs\footnote{Cardboard SDK~\cite{website:cardboard-sdk}, Google VR SDK~\cite{website:google-vr-sdk}, VIVE SDK~\cite{website:vive-sense-sdk}, Magicverse SDK~\cite{website:magicverse-sdk}, and UtoVR SDK~\cite{website:utovr-sdk}.}.}
\textblue{(2) Because no equivalent effective iOS library-identification method was available~\cite{paper:harmful-ios-lib-detection}, we matched the identified Android applications to iOS versions by application title and developer. We then collected the available English reviews for all included applications.}
\textblue{This sampling frame covers the public English reviews that were accessible from the selected stores and filtering procedures; it should not be interpreted as a complete census of all VR applications, all VR users, or all review behavior across marketplaces.}
\textred{Table~\ref{tab:app_platforms} reports the observed review-date range for each marketplace when timestamps were available. These ranges describe the collected records, not the complete operating history of each store.}
\begin{table}[t!]
\centering
\renewcommand{\arraystretch}{1.5} %
\resizebox{\columnwidth}{!}{%
\begin{tabular}{|c|c|}
\hline
\textred{\textbf{Marketplace}} & \textred{\textbf{Observed Review-Date Range}} \\
\hline
\textred{Meta Horizon Store} & \textred{2021-02-15 / 2025-06-15} \\
\hline
\textred{VIVEPORT} & \textred{2016-04-15 / 2024-03-02} \\
\hline
\textred{SideQuest} & \textred{2019-09-11 / 2025-06-11} \\
\hline
\textred{Meta Horizon App Lab} & \textred{2019-05-20 / 2025-06-10} \\
\hline
\textred{Steam} & \textred{2015-05-30 / 2025-07-07} \\
\hline
\textred{Google Play} & \textred{2013-04-12 / 2025-07-12} \\
\hline
\textred{iOS App Store} & \textred{Unavailable from the collected public metadata} \\
\hline
\end{tabular}%
}
\caption{\textred{Observed Review Timespans by Marketplace}}
\label{tab:app_platforms}
\end{table}

\subsection{Preprocessing}

\label{subsubsec:methodology-preprocessing}
User reviews
are often unstructured texts containing misspelled, informal, and spam expressions~\cite{paper:app-love-users-ase15, paper:oasis-fse17, paper:idea-online-app-review-icse18}.
To facilitate the analysis, we follow the typical preprocessing steps from literature~\cite{paper:change-requests-icse17, paper:keyatm-icse22} to process the raw data before taxonomy construction:
(1) basic data cleaning and format normalization,
(2) contraction and abbreviation expansion,
(3) spelling correction,
(4) tokenization,
(5) lemmatization,
(6) stop word removal,
and (7) redundancy reduction. 
The details of implementation can be found on our website.

In our dataset, many reviews contain multiple sentences on various topics. 
For example, ``\textit{The pros and cons of this application: (1) Innovative locomotion scheme...; (2) Great performance...; (3) Does not have a tactile interface, hand passes right through surfaces...}''.
Hence, we split one single review into a list of sentences to ease the analysis process, resulting in a total of 3,372,157 review sentences for analysis.
The subsequent analyzing approaches are conducted on the sentence-level granularity.

\begin{table*}[t!]
	\centering
	\caption{\textblue{VR App Review Dataset}}
    \vspace{-0.8em}
	\label{table:xrzoo}
	\resizebox{\textwidth}{!}{
	\begin{threeparttable}
	\begin{tabular}{l l r r}
		\toprule
		\textblue{\textbf{Marketplace}} &
        \multicolumn{1}{c}{\begin{tabular}[c]{@{}c@{}}\textblue{\textbf{Device}}\\\textblue{\textbf{Class(es)\tnote{1}}}\end{tabular}} &
        \multicolumn{1}{c}{\begin{tabular}[c]{@{}c@{}}\textblue{\textbf{\# of Apps}}\\ \textblue{\textbf{Crawled}}\end{tabular}} &
        \multicolumn{1}{c}{\begin{tabular}[c]{@{}c@{}}\textblue{\textbf{\# of Reviews}}\\ \textblue{\textbf{Crawled}}\end{tabular}} \\
		\cmidrule(){1-4}
		\textblue{Meta Horizon Store~\cite{website:oculus-app-store}} & \textblue{S, P, M} & \textblue{9,848} & \textblue{445,720} \\
        \cdashlinelr{2-4}
		\textblue{VIVEPORT~\cite{website:viveport}} & \textblue{S, P} & \textblue{3,147} & \textblue{12,597} \\
        \cdashlinelr{2-4}
		\textblue{SideQuest~\cite{website:sidequest}} & \textblue{S, P, M} & \textblue{6,964} & \textblue{49,696} \\
        \cdashlinelr{2-4}
		\textblue{Meta Horizon App Lab~\cite{website:oculus-app-lab}} & \textblue{S} & \textblue{2,294} & \textblue{57,445} \\
		\cdashlinelr{1-4}
		\textblue{Steam~\cite{website:steam-app-store-vr}} & \textblue{P} & \textblue{5,966} & \textblue{559,872} \\
		\cdashlinelr{1-4}
		\textblue{Google Play~\cite{website:google-play}} & \textblue{M} & \textblue{360} & \textblue{522,792} \\
        \cdashlinelr{2-4}
		\textblue{iOS App Store~\cite{website:apple-app-store}} & \textblue{M} & \textblue{166} & \textblue{8,846} \\
		\cmidrule(){1-4}
        \textblue{Total} & \textblue{S, P, M} & \textblue{28,745} & \textblue{1,656,968} \\
        \bottomrule
	\end{tabular}
    \begin{tablenotes}
        \item[1] \textblue{S, P, and M stand for applications running on standalone, PC-assisted, and mobile device sets, respectively. We report only device classes and counts verified from our collection metadata; we do not infer marketplace-wide operating systems, SDKs, or supported interactions for all apps.}
	    \item[] \textblue{Counts reflect the revised crawl summarized by the observed review-time ranges in Table~\ref{tab:app_platforms}.}
    \end{tablenotes}
    \end{threeparttable}
    }
\end{table*}

\subsection{Taxonomy Construction}
\label{subsec:taxonomy-construction}
\textblue{In this section, we aim at building taxonomies of the quality attributes}
\textblue{and the factors that influence user experience on these quality attributes.}
\textblue{To perform an efficient and reliable review analysis on the massive dataset,}
\textblue{we design a semiautomatic procedure that uses automatic NLP methods for candidate generation and manual qualitative coding for conceptual interpretation.}
\textblue{Specifically, we first perform unsupervised aspect extraction to automatically extract candidate tokens that represent key aspects commonly discussed in the reviews (e.g., ``\textit{locomotion}'', ``\textit{haptics}'', ``\textit{comfort}''). We then manually inspect these candidates with their surrounding review contexts and construct the two taxonomies based on the reviewed concepts.}
\textblue{The correlations between the quality attributes and influencing factors are analyzed further in Section~\ref{subsec:method-manual-analysis}.}

\subsubsection{Automatic Aspect Extraction}\label{subsec:methodology-aspect-extraction}
Unsupervised aspect extraction methods are widely used to distill key topics from massive textual data~\cite{paper:lda-topic-modelling-survey, paper:aspect-extraction-survey}.
To capture aspects as many as possible for facilitating the subsequent manual taxonomy construction process, we adopt multiple popular extraction methods and combine their results, instead of using a single approach.

Guided by the systematic literature review procedure discussed by Kitchenham and Charters~\cite{paper:kitchenham2007guidelines}, we perform literature searching and selection to choose aspect extraction methods.
During literature searching, we use keywords (\textit{aspect extraction}, \textit{topic model}, \textit{review analysis}) to find related 
papers published in top venues of both SE and NLP fields\footnote{SE venues: ICSE, ESEC/FSE, ASE, ISSTA, TSE, TOSEM; NLP (AI) venues: ACL, NeurIPS, ICML, EMNLP, NAACL, COLING.} and arXiv (for both SE \& NLP) during the last six years.
After that, we select aspect extraction methods according to the evaluation results in the papers, the number of paper citations and GitHub stars of the tools, we choose both 
(1) the latest state-of-the-art methods (we choose two methods with the state-of-the-art evaluation results in SE and NLP fields respectively), and 
(2) the classic methods with rather accurate and stable performance, which are recognized by the community (we choose the Top-3 methods with the highest sum of standard scores (Z-score) of the paper citation number (C) and GitHub stars of the tools (S)).
The sum of Z-scores is calculated to avoid being dominated by one field:
\begin{gather}
Z_{sum} = \frac{C - \mu_{C}}{\sigma_{C}} + \frac{S - \mu_{S}}{\sigma_{S}}
\end{gather}
where \( \mu \) and \( \sigma \) represent the mean value and the standard deviation of C and S among all methods, respectively.

Based on the criteria, we select the following methods:
Hierarchical Dirichlet Processes (HDP~\cite{paper:hdp, paper:change-requests-icse17, paper:where2change-tse21}), Biterm Topic Model (BTM~\cite{paper:btm, paper:emerging-issue-detection-tse21}), Keyword Assisted Topic Models (keyATM~\cite{paper:keyatm-arxiv20, paper:keyatm-icse22}), Attention-based Aspect Extraction (ABAE~\cite{paper:neural-ae-acl17}), and BERTopic (~\cite{paper:bertopic}). 

\begin{table}[t]
	\centering
	\caption{\textblue{Aspect Extraction Inputs to Open Coding}}
	\scriptsize
	\label{table:aspect-extraction-approaches}
	\resizebox{\linewidth}{!}{
	\begin{threeparttable}
	\begin{tabular}{L{3.1cm}L{3.5cm}R{1.15cm}}
		\toprule
		\textblue{\textbf{Method}} & \textblue{\textbf{Candidate representation}} & \textblue{\textbf{NPMI}} \\ \midrule
		\textblue{Hierarchical Dirichlet Processes (HDP)~\cite{paper:hdp}} & \textblue{Topic--term groups} & \textblue{0.045} \\
		\rowcolor{mygray}
		\textblue{Biterm Topic Model (BTM)~\cite{paper:btm}} & \textblue{Topic--term groups} & \textblue{0.083} \\
		\textblue{Keyword Assisted Topic Models (keyATM)~\cite{paper:keyatm-arxiv20}} & \textblue{Keyword-assisted topic--term groups} & \textblue{0.129} \\
		\rowcolor{mygray}
		\textblue{Attention-based Aspect Extraction (ABAE)~\cite{paper:neural-ae-acl17}} & \textblue{Aspect groups with representative terms} & \textblue{0.166} \\
		\textblue{BERTopic~\cite{paper:bertopic}} & \textblue{Document clusters with topic terms} & \textblue{0.163} \\ \bottomrule
	\end{tabular}
	\begin{tablenotes}
		\item[] \textblue{NPMI values are the final coherence scores reported in the public study artifact.}
	\end{tablenotes}
	\end{threeparttable}
	}
\end{table}

Multiple parameters can be tuned for the aspect extraction methods (e.g., the number of aspect groups).
Following the literature~\cite{paper:keyatm-icse22, paper:bertopic, paper:topic-coherence-eacl14}, we utilize normalized pointwise mutual information (NPMI)~\cite{paper:npmi} as the metric to assist parameter tuning:
\begin{gather}
\operatorname{NPMI}\left(w_{i}, w_{j}\right)=\frac{\log \frac{p\left(w_{i}, w_{j}\right)}{p\left(w_{i}\right) p\left(w_{j}\right)}}{-\log p\left(w_{i}, w_{j}\right)},
\end{gather}
where $p\left(w_{i}\right)$ and $p\left(w_{j}\right)$ represent the probability of two aspect terms $w_{i}$ and $w_{j}$ in the whole corpus, respectively. 
When $w_{i}$ and $w_{j}$ are independent, $p\left(w_{i}\right) p\left(w_{j}\right)$ will be the joint probability of word pair. 
The actual probability of their co-occurrence is $p\left(w_{i}, w_{j}\right)$.

NPMI measures the coherence of topics, which evaluates the semantic consistency between the aspect terms in a group~\cite{paper:topic-coherence-meaning-conll12}. 
The higher the NPMI scores are, the more semantically coherent the grouped aspects will be.
To ensure the quality and interpretability of the produced aspect groups,
\textblue{we tune the parameters for each method using NPMI as a diagnostic of topic coherence. Table~\ref{table:aspect-extraction-approaches} reports the final scores: 0.045 for HDP, 0.083 for BTM, 0.129 for keyATM, 0.166 for ABAE, and 0.163 for BERTopic. NPMI supports parameter selection, but it is not automated proof that an extracted term is a valid quality attribute or influencing factor. Exact per-tool candidate counts before normalization are unavailable because the original intermediate snapshots were not retained; we therefore do not reconstruct them retrospectively.}
\textblue{Finally, we combine the candidate aspect terms produced by these five methods for the subsequent manual taxonomy construction by merging the method outputs into one candidate list and removing exact duplicates where applicable. During open coding, evaluators inspect candidate terms together with their review contexts; we do not treat tool-specific clusters or intermediate outputs as final taxonomy categories, and we do not require agreement among all five extraction methods before a candidate can be inspected.}

\subsubsection{Manual Taxonomy Construction}
\label{subsec:method-taxonomy-construction}
In this section, we aim at manually inspecting the extracted aspects described in Section~\ref{subsec:methodology-aspect-extraction} to build the taxonomies of quality attributes and influencing factors, with the process illustrated in Figure~\ref{fig:overview}.

\textblue{We follow the widely-used open coding procedure~\cite{book:open-coding-16} for the manual inspection, which involves five evaluators who all have more than five years of software development experience and sufficient knowledge on VR.}
During open coding, researchers usually attach \textit{conceptual labels} to the \textit{data} under analysis, resulting in a \textit{taxonomy
of characterizing categories}~\cite{book:open-coding-16}.
\textblue{In our work, we use the automatically extracted candidate aspects as \textit{data} under analysis. For each candidate, evaluators inspect surrounding review sentences and local usage contexts, decide whether the candidate reflects a quality attribute, an influencing factor, or neither, and then attach quality attributes and influencing factors of hierarchical structures as \textit{conceptual labels}.}

\textblue{The five evaluators acted as qualitative coders rather than independent raters of a fixed label set. The initial taxonomies are constructed through an extensive scanning of all aspects, during which we group relevant aspects and abstract the core concepts. In each of the following iterations, we scan all aspects by adding or updating labels that are more appropriate and accurate to some specific aspects, merging similar labels, removing inadequate labels, adjusting hierarchical structures, refining unclear definitions, and updating a shared coding guide with category definitions, inclusion/exclusion rules, examples, and boundary cases.}
\textblue{For determining the taxonomy of software quality attributes of VR applications, we also use the ISO standard of system and software quality model~\cite{standard:iso-25010-square-2011} as a sensitizing reference for common software-quality terminology rather than as a closed coding schema. VR-specific categories are added when review evidence does not fit the ISO model.}
\textblue{During each iteration, newer versions of the two taxonomies are produced and evaluated by all the evaluators. Disagreements are resolved through discussion, guide revision, and reinspection of review contexts until consensus, rather than by majority vote or by calculating inter-rater reliability over a fixed codebook. The iteration stops only if consensus is reached among the evaluators and the evaluators cannot make a further refinement.}

\textblue{In this work, the manual inspection goes through five iterations which took about two months to complete,}
\textblue{ending up with two \textit{taxonomies of characterizing categories} regarding quality attributes and influencing factors, respectively.}

\subsection{Focused Coding and Quantitative Analysis}
\label{subsec:method-manual-analysis}
In this stage, we manually analyze a representative subset of user reviews based on the results of Section~\ref{subsec:method-taxonomy-construction}, to validate and refine our taxonomies~\cite{book:sage-qda-focused-coding}.
To achieve a 95\% confidence level and ±5\% precision with the total amount of 3,372,157 segmented review sentences, the sample size has to be no less than 385.
To make the results more reliable, we randomly sample \reviewSampleSize review sentences.
\textblue{The \reviewSampleSize-sentence evaluation set was constructed by stratified random sampling over app-store strata, using the app-store distribution of the collected corpus as the sampling target. This sampling step supports analysis of the collected review corpus; it does not claim representativeness over users whose reviews are unavailable, non-English reviews, or marketplaces outside the collection frame.}

We follow the diluted focused coding procedure~\cite{book:sage-qda-focused-coding} and perform in iterations. 
During focused coding, we compare and merge similar categories, remove inadequate categories, adjust hierarchical structures,
refine unclear definitions, etc.
Considering that the analyzers involved in the taxonomy construction process in Section~\ref{subsec:method-taxonomy-construction} may bring bias to the quantitative analysis,
we invite two new analyzers and only one previously participated analyzer for the focused coding process.
Both new analyzers have over three years of software development experience and are familiar with VR.

Before the first iteration, the previously participated analyzer explains the definitions of the taxonomy constructed in Section~\ref{subsec:method-taxonomy-construction} to the new analyzers.
Next, all analyzers independently label each sampled review sentence with the quality attribute(s) and influencing factor(s) mentioned in it, and then discuss the discrepancies in results, form a new version of taxonomies, and reanalyze\footnote{Analysis guidelines for analyzers can be found on our website.}.
\textblue{In the first pass of each focused-coding round, the three analyzers label independently using the current coding guide. The available study records do not document a separate pilot outside the four reported rounds; the first round also provided calibration through comparison and discussion. The analyzers then compare labels, discuss disagreements at the sentence and category-boundary levels, revise definitions where necessary, and relabel affected samples in the next round. Because the label definitions are intentionally refined across rounds, the process is an evolving consensus-coding design rather than a fixed-rating reliability study; therefore, we do not report IRR statistics such as Cohen's or Fleiss' kappa. The two new analyzers reduce carry-over bias from taxonomy construction, while retaining one experienced coder keeps the definitions consistent with the open-coding results. This design mitigates, but cannot eliminate, experienced-coder bias.}
The quality attributes and influencing factors are labeled using the taxonomies from Section~\ref{subsec:method-taxonomy-construction} in the first iteration, and then labeled using the latest taxonomies from the last iteration in the subsequent iterations.
After seven weeks of analysis and \manualAnalysisIterationNum iterations of focused coding, the three analyzers achieve a consensus.
\textred{As a supplementary analysis requested by the reviewers, we additionally sampled \updatedReviewSampleAmount review sentences from the revised crawl while preserving the corpus's app-store proportions. This updated-corpus sample supports the platform- and genre-based analyses; it is separate from the original 7,000-sentence human-coded dataset.}

\textred{
Labels for quality attributes and influencing factors in the \updatedReviewSampleAmount-sentence updated-corpus sample were assigned in two stages. First, three large language models (DeepSeek, Qwen, and Zhipu) were prompted via their APIs to annotate each sentence; only samples receiving identical labels from at least two models were retained for manual review. Second, all retained labels were manually reviewed to correct residual model errors and finalize the annotations. The temporal analysis uses a separate sample of \temporalSamplePerQuarter reviews in each of 20 quarters, and the accessibility analysis uses a separately retrieved and human-screened subset of 675 reviews. Update-log labels are produced from a separate developer-log corpus, as described in Section~V-K. These supplementary pipelines were not used to construct the original taxonomies or original manually labeled dataset, and they should not be interpreted as detecting manipulated or non-genuine reviews.
}
Note that the sampled dataset contains review sentences unrelated to software quality, e.g., ``\textit{During Steam sales, pick up all Batman games for \$5, get the whole series for \$30}''.
We remove all such sentences and finally obtain a labeled dataset containing \informativeReviewAmount review sentences.
We then use the labeled dataset
for a deeper quantitative analysis.

To further explore the factors that influence user experience regarding each quality attribute,
we manually examine the reviews labeled with both this quality attribute and different influencing factors, with the assumption that: (1) one review labeled both the quality attribute and influencing factor indicates a possible correlation between them, and (2) more such reviews indicate a stronger correlation.

\begin{figure}[t!] 
	\centering 
	\includegraphics[width=\columnwidth,trim={0 9.95cm 0 0},clip]{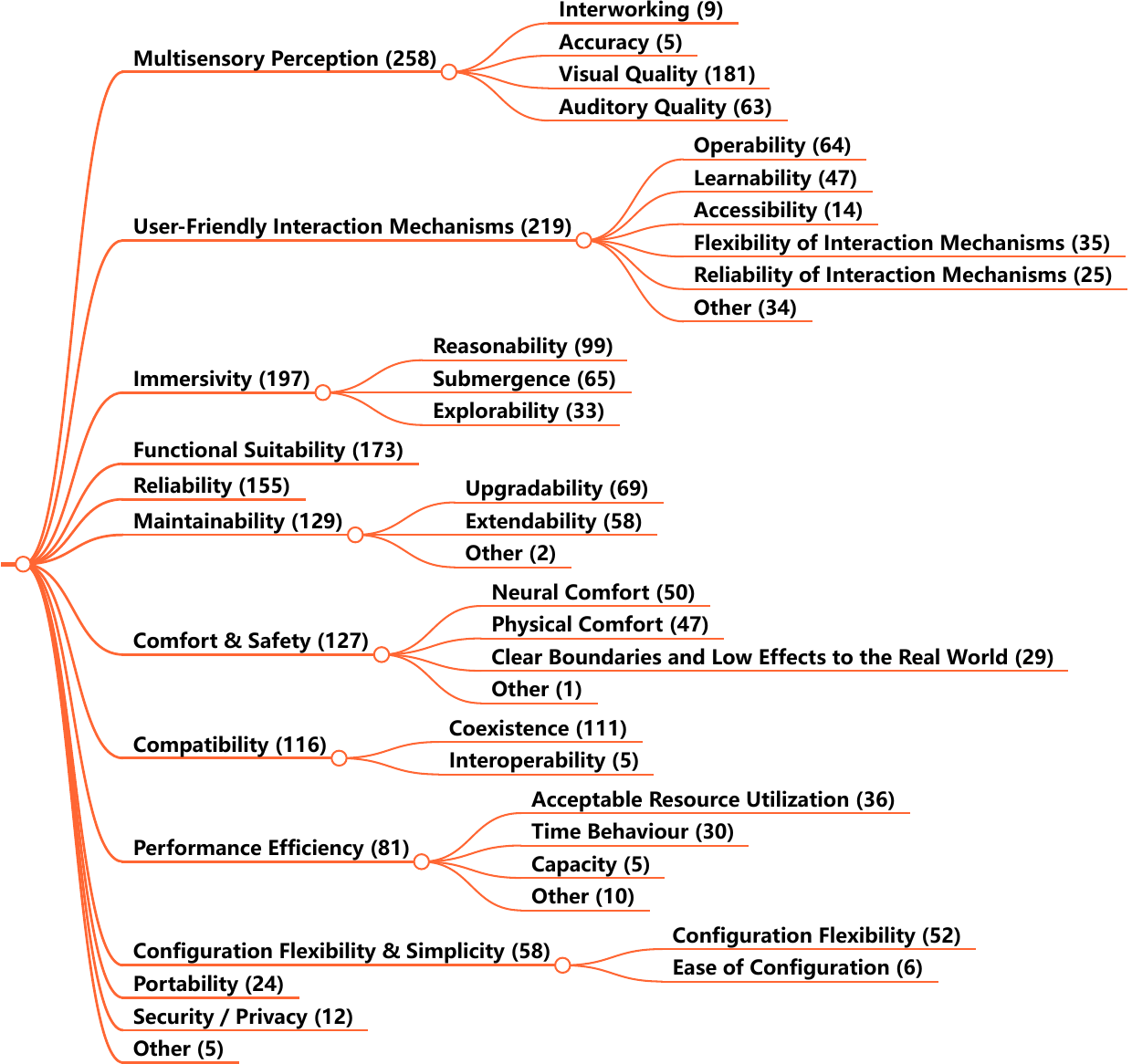} 
	\caption{The Taxonomy of Quality Attributes}
	\label{fig:quality-taxonomy}
\end{figure}

\begin{figure}[t!] 
	\centering 
	\includegraphics[width=0.95\columnwidth,trim={0 4cm 0 0},clip]{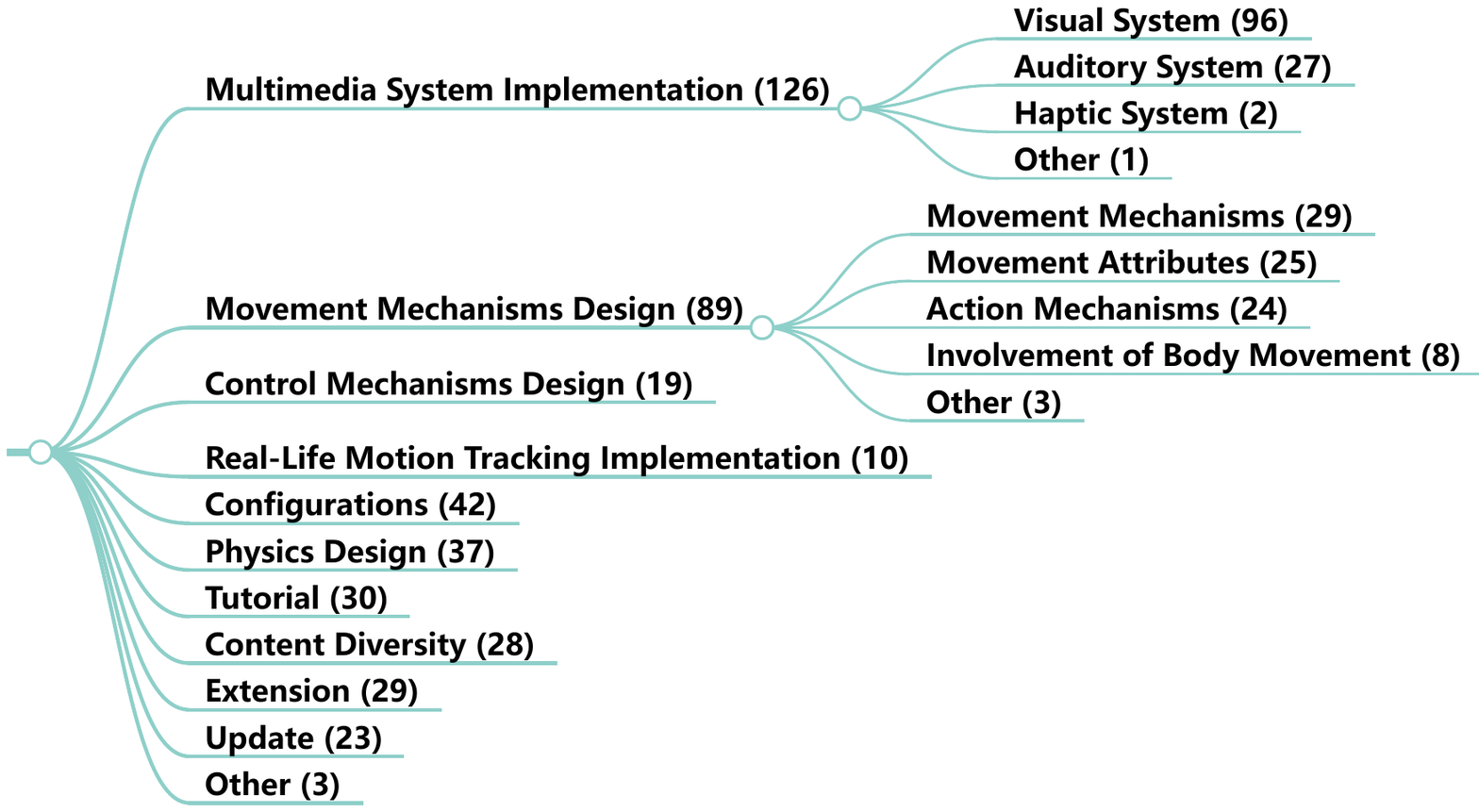} 
	\caption{The Taxonomy of Influencing Factors\protect\footnotemark}
	\label{fig:factor-taxonomy}
\end{figure}

\footnotetext{Some reviews do not mention influencing factors.}

\begin{table*}[t!]
\centering
\scriptsize
\caption{Descriptions and User Review Examples of the Root-Level Quality Attributes (QAs)}
\label{table:review-example}
\resizebox{\textwidth}{!}{%
\begin{tabular}{L{1.5cm}L{8.6cm}L{7.25cm}}
\toprule
\textbf{QA} & \textbf{Description} & \textbf{User Review Example of (Un)satisfactory Experience on the QA} \\ \midrule
Multisensory Perception & VR apps should make multimodal behavior of multiple senses (visual, auditory, haptic, etc.) interwork with each other, and make monomodal behavior of every single sense accurate and of high quality. & (Unsynchronized effects between graphics and audio) ``\textit{I do not enjoy this game as the sounds are not synchronized to actions.}'' \\
\cmidrule(lr){1-3}
UF Interaction Mechanisms & VR apps should be easy to interact with and be used normally by a wide range of people with special needs. The interaction mechanisms should be flexible, reliable, and easily learned.  &  (Poor accessibility for limited mobility) ``\textit{I must play from a sitting position due to injury. This game makes me painfully reach either too high or too low to comfortably play.}''  \\
\cmidrule(lr){1-3}
Immersivity & VR apps should make users feel detached from real life and submerge in a coherent and reasonable virtual world, where they can interact and explore in various dimensions. & (Poor submergence) ``\textit{This app restricts me from moving around freely, which is a huge immersion breaker. If I wanna step forward, I have to press a button instead of just walking.}''  \\
\cmidrule(lr){1-3}
Functional Suitability & VR apps should offer functionalities that align with both explicitly stated and implicitly inferred user needs when operated under designated conditions. This ensures that the software not only meets the direct requirements but also anticipates potential user expectations. & \textit{"Most videos are rotated 90° to the right and can't be recentered using the home button."} \\
\cmidrule(lr){1-3}
Reliability & VR apps should consistently perform their specified functions under delineated conditions over a predetermined duration. This consistency in performance is pivotal for user trust and satisfaction, ensuring that the virtual experience remains uninterrupted and dependable. & (Bad reliability) "\textit{The game just sends me to a black screen.}" \\
\cmidrule(lr){1-3}
Maintainability & VR apps should be designed such that they can be efficiently and effectively modified by the designated maintenance teams. This ensures that the software can evolve, adapt, and be rectified with minimal effort, thereby prolonging its lifecycle and relevance. & (Inferred well maintainability) \textit{"This game appears to be pretty consistently updated and improved."}  \\
\cmidrule(lr){1-3}
Comfort \& Safety & VR apps should be both neurally and physically comfortable to users. & (Neural discomfort) ``\textit{The teleportation pulls me quickly in a sprint motion to the target spot, causing lots of nausea.}''  \\
\cmidrule(lr){1-3}
Compatibility & VR apps should make multiple components or systems supporting VR apps able to coexist in a same environment and share resources with each other without harm. & (Incompatible controllers) ``\textit{The game isn't compatible with Valve Index controllers. It is impossible to change the weapon and the buttons are not properly assigned.}''  \\
\cmidrule(lr){1-3}
Performance Efficiency & VR apps should be efficient in processing and responding, the maximum constraints of performance should match requirements, and the resource utilization should be acceptable. & (Bad performance) ``\textit{I've got the latest drivers i7@4.2, 1080TI, and the game is installed on an SSD, yet I'm getting dropped frames.}''  \\
\cmidrule(lr){1-3}
Configuration Flexibility \& Simplicity & The configuration of VR apps should be flexible and easy for users to configure. &  (Inflexible configuration) ``\textit{A shortcoming to this game is that you can only choose between three graphics presets.}'' \\
\cmidrule(lr){1-3}
Portability & VR apps should be constructed to allow for a seamless transfer between different hardware, software, or other operational environments. This quality ensures that users can experience the virtual realm across various platforms without compromising on functionality or experience. & (Poor installability) \textit{"Failed to install. I've tried installing multiple times."} \\
\cmidrule(lr){1-3}
Security / Privacy & VR apps should ensure users' security and privacy during normal function operation, content sharing, etc. & (Insecure private room) ``\textit{I wish their passwords could be more complex. We have had people jump into our private room family games before, not as fun}.'' \\ 
\bottomrule
\end{tabular}%
}
\end{table*}

\begin{table*}[t!]
	\centering
	\caption{User Review Examples of Important Quality Attributes (QAs)}
	\label{table:qa-review-example}
	\footnotesize
	\begin{threeparttable}
	\begin{tabular}{llL{4.5cm}L{9.5cm}}
		\toprule
		\textbf{QA} & \textbf{Sub QA} & \textbf{Description} & \textbf{User Review Example$^{1}$ of Unsatisfactory Experience on the Quality Attribute} \\ 
		\cmidrule(){1-4}
		\multirow{3}{*}{\rotatebox[origin=c]{90}{Multisensory Perception}} & Interworking & Multisensory behavior for different senses should interwork with each other and app logic. & (Unsynchronized effects between graphics and audio) ``\textit{I do not enjoy this game as the sounds are not synchronized to actions.}'' \\
		\cmidrule(lr){2-4}
		& Accuracy & Sensory effects (i.e., visual, auditory, or haptic) should be accurate. & (Inaccurate audio effects) ``\textit{The positional audio is totally wrong in many cases. Audio that should be in front of you can be positioned far above you or even behind you. Please fix this!}'' \\
		\cmidrule(lr){2-4}
		& \multirow{1}{*}{Visual Quality} & Quality of visual effects (graphic clarity, contrast, smoothness, visual fidelity...) & (Unsmooth scene transitions) ``\textit{The experience isn't good, I get motion sickness because of the stuttering effects during cart riding scenes.''} \\
		\cmidrule(lr){2-4}
		& Auditory Quality & Quality of auditory behavior (3D spatial effects, audio clarity, audio fidelity...) & (Low quality on 3D spatial effects) ``\textit{This app missed the opportunity to make good use of positional audio. This is basically no different from a 360\degree~video.}'' \\
		\cmidrule(lr){1-4}
		\multirow{5}{*}{\multirow{3}{*}{\rotatebox[origin=c]{90}{User-Friendly Interaction Mechanisms}}} & Operability & Apps should be easy to operate and interact with. & ``\textit{My big complaint is that it is much harder to do what you want to do. I would find myself trying to execute a certain move a certain way and even after lots of practice it was still hard to pull off certain moves.}'' \\
		\cmidrule(lr){2-4}
		& Learnability & The interaction mechanisms should be easily learned or inferred. & ``\textit{Contrary to what you think intuitively, you do not need to swing your arms to launch yourself through the air. This isn't explicitly stated, so I spent my first five hours in the game swinging my arms around like an idiot before I discovered you just have to point and click.}'' \\
		\cmidrule(lr){2-4}
		& Accessibility & Apps can be used normally by a wide range of people in different situations. & ``\textit{I must play from a sitting position, due to significant injury. The game makes me painfully reaching either too high, or too low to comfortably play. I speculate there were few mobility-limited testers involved.}'' \\
		\cmidrule(lr){2-4}
		& \begin{tabular}[l]{@{}l@{}}Flexibility of \\ Interaction Mech.\end{tabular} & The interaction mechanisms shouldn't be too restricted. & ``\textit{Lacking of free movement displeases me greatly, doesn't have any flexibility for people who prefer free movement or smooth turning.}'' \\
		& & & (Inflexibility of teleportation) ``\textit{The main problem with the game is the reliance upon teleportation. We can't move while shooting a gun. Why can't just run by pushing the touchpad forward?}'' \\
		\cmidrule(lr){2-4}
		& \begin{tabular}[l]{@{}l@{}}Reliability of \\ Interaction Mech.\end{tabular} & The interactions should be stable in process and precise in results. & ``\textit{The interaction is imprecise and wonky to an immersion-breaking degree.}'' \\
		& & & (Imprecise tracking system) ``\textit{BAD THINGS - Head/controller tracking sometimes not tracking correctly (my arms fly 20 meters around me for the entire game).}'' \\
		\cmidrule(lr){1-4}
		\multirow{3}{*}{\multirow{3}{*}{\rotatebox[origin=c]{90}{Immersivity}}} & Reasonability & The functionalities and logic should be reasonable and coherent. & (Imprecise tracking) ``\textit{The gun angle is tilted upwards, so your hands in VR and your hands in real life don't match up, killing any immersion that could be salvaged with this game.}'' \\
		& & & (Bad physics implementation) ``\textit{physics don't work well, for example, I hit the ball when it was close to me and instead of going forward, it goes up or doesn't move at all.}'' \\
		\cmidrule(lr){2-4}
		& Submergence & Make users feel immerged/surrounded by virtual worlds and detached from real life. & ``\textit{There are restrictions put in place that stop me from moving around freely in real life, this is a huge immersion breaker. You're telling me if I wanna step forward I have to press a button instead of just walking?}'' \\
		\cmidrule(lr){2-4}
		& Explorability & Make users able to explore in different combinations of routes and dimensions. & ``\textit{This game has less explorability and is much more linier than its three predecessors.}'' \\
		\cmidrule(lr){1-4}
		\multirow{1}{*}{\multirow{1}{*}{\rotatebox[origin=c]{90}{Comfort}}} & Neural Comfort & Make users feel neurally comfortable. & ``\textit{It is incredibly nauseating to have your head stop in the game but continue to move in the real world.}'' \\
		& & & (Poor virtual movement system design) ``\textit{One thing wrong with this game is that teleportation is awful because it pulls you quickly in a sprint motion to the spot you are teleporting, which causes me lots of nausea.}'' \\
		\cmidrule(lr){2-4}
		& Physical Comfort & Make users feel physically comfortable. & ``\textit{Holding down the two grip buttons was making my hands cramp.}'' \\
		\bottomrule
	\end{tabular}%
    \begin{tablenotes}
		\item[1] The review examples in this paper are original ones with some cosmetic modifications or grammar corrections. The raw reviews can be found on our website.
    \end{tablenotes}
    \end{threeparttable}
\end{table*}

\section{Findings and Implications}
\label{sec:answers-to-rqs}

\subsection{Results Overview}

Figures~\ref{fig:quality-taxonomy} and~\ref{fig:factor-taxonomy} show the full taxonomies of quality attributes and influencing factors, respectively, along with the number of review sentences classified into each category.
Table~\ref{table:review-example} shows the descriptions and user review examples of the most specific quality attributes (QAs) to VR applications.

As demonstrated in Figure~\ref{fig:quality-taxonomy}, among the \qualityFirstLayerCategoryAmount software quality attributes concerned by VR application users,
we find that: 

(1) Some of the quality attributes are literally similar to those of the common software according to the ISO software quality model~\cite{standard:iso-25010-square-2011}, but they may involve different interpretations (e.g., compatibility, performance efficiency, security \& privacy, etc.).
For example, although compatibility is already a hard problem for other software, VR developers encounter greater challenges as there are more components that developers need to ensure the capability of coexistence.
Specifically, developers may need to ensure that their software supports all mainstream HMDs, controllers, advanced VR devices, dedicated VR devices, GPUs, operating systems, etc.

(2) Some of the quality attributes are more specific to VR applications, denoted as \textit{VR-specific quality attributes}, including multisensory perception, user-friendly interaction mechanisms, immersivity, and comfort \& safety.
These quality attributes are closely related to the unique properties of
VR applications, e.g., revolutionized interaction mechanisms and immersive experiences.
For example, due to the inconsistent movement states between immersive virtual worlds and the real world, lots of users experience and complain about 
discomfort in VR applications, which are rarely observed in traditional software. 
Such issues bring severe consequences to users’ health and safety.
We observe that the VR-specific quality attributes are discussed in 51.5\% of the labeled reviews, indicating a great concern by VR users.
Table~\ref{table:qa-review-example} demonstrates the descriptions of these quality attributes with user review examples in our dataset.

\stepcounter{subfidcounter}
\begin{findingbox}
	\begin{finding}
		There exist \qualityFirstLayerCategoryAmount quality attributes of major concern to VR application users. Some quality attributes are more specific to VR applications, especially
       \textbf{multisensory perception, user-friendly interaction mechanisms, immersivity, and comfort \& safety}. 
	\end{finding}
\end{findingbox}
As shown in Figure~\ref{fig:factor-taxonomy}, we find \factorFirstLayerCategoryAmount major factors that can influence the software quality of VR applications.
The taxonomy of influencing factors and the correlations between these factors and quality attributes (further introduced in the latter part of Section~\ref{sec:answers-to-rqs}) can suggest components to evaluate for each quality attribute and provide prospective directions for improving user experience on each attribute.
From the significant amount of complaints raised by users on VR-specific quality attributes, we find that the reported issues are often caused by bad design or implementation of movement mechanisms, control mechanisms, real-life motion tracking, multimedia system, physics and application content.
Most of them are special in VR applications compared with traditional software.

Due to the large proportion of reviews discussing VR-specific quality attributes (51.5\%),
these quality attributes tend to bring new challenges to the developers and researchers of VR applications, but have not been comprehensively studied by previous work. Therefore, we focus on analyzing the four VR-specific quality attributes in the following sections. Specifically, for each of the four quality attributes, we present the detailed user concerns and influencing factors, and discuss implications for both developers and researchers.
The descriptions and review examples of other quality attributes and influencing factors can be found on our website.

\subsection{Multisensory Perception}
\label{subsec:rqs-multisensory}

\subsubsection{\textbf{Detailed User Concerns}}
\label{subsubsec:rqs-multisensory-userconcerns}

Traditional software delivers information to users through a restricted screen-sized interface and audio, where the types and amount of information conveyed are limited.
However, VR applications convey information 
to users through multiple channels of perception (e.g., visual, auditory, and haptic senses, etc.) with fewer restrictions, to offer a better sense of immersion and interaction experiences.
This makes multisensory perception the most concerned quality attribute by users, which is discussed in more than 16.6\% (258/\informativeReviewAmount) of the labeled reviews.

\textbf{(a) The sub quality attributes related to multiple senses:}

\textbf{Interworking}.
The digital simulation for different kinds of perception needs to interwork with each other,
and performs synchronously and congruously.
According to related user reviews, poor interworking issues are often caused by unsynchronized effects or mismatched information provided by different senses.
For example, one user raised a complaint about unsynchronized effects between graphics, audio, and haptic effects: ``\textit{The vibration of controllers and the audio are not synchronized to actions at all}''.
Such issues usually break users' immersion and make them confused about how to interact with the apps.

\textbf{Accuracy}.
VR applications need to make each of their sensory effects 
accurate.
Incorrect information conveyed can confuse or even mislead users.
For example, one user complained that the audio effects are inaccurate: 
``\textit{Audio that should be in front of you can be positioned far above you or even behind you.}''
As VR applications often use spacial audio to tell which direction the interactable objects or unexplored scenes lie in, such misinformation can lead users to wrong directions in the virtual environments.

\textbf{(b) The sub quality attributes related to a single sense:}

\textbf{Visual (Graphic) Quality}.
Users' complaints about visual quality mainly include smoothness (during scene transition or user movement), visual fidelity (how accurate the displayed results are), graphic clarity, contrast, etc.
Low visual quality (e.g., unsmooth scene transitions) hurts user experience and even makes users feel uncomfortable.
For instance, 
one user expressed frustration over unsmooth scene transitions inducing dizziness, noting,
``\textit{I get motion sickness because of the stuttering effects during cart riding scenes. It's not laggy, just not smooth''}.

\textbf{Auditory Quality}.
VR applications support richer capabilities than traditional software to deliver information through audio.
Users mainly expect to experience changes in acoustic intensity and orientation when they change their position or orientation in VR worlds, and care much about the acoustic effects of VR apps.
In our dataset, the major complaints from users lie in 3D spacial audio effects, audio clarity, audio fidelity (how accurately the played soundtracks reproduce their original ones), etc.
For example, a user commented on an application of low quality on 3D spatial audio effects that ``\textit{This application doesn't use positional audio. It is basically no different from a 360\degree~video.}''
Such low auditory quality makes users difficult to interact with VR applications in stereoscopic environments.

\stepcounter{subfidcounter}
\begin{findingbox}
	\begin{finding}
       Multisensory perception is the most concerned quality attribute by users. 
       Users expect (1) the interworking of 
       multiple senses (visual, auditory, haptic perceptions, etc.), and (2) 
       high-quality effects
       of every single sense.
		Users have high expectations on the quality of visual and auditory effects, e.g., 3D spacial audio effects, for VR applications.
	\end{finding}
\end{findingbox}

\subsubsection{\textbf{Major Influencing Factor}}
\label{subsubsec:rqs-multisensory-factors}

\textbf{Multimedia System Implementation.} Poor multisensory perception is usually caused by the bad implementation of the multimedia system.
The multimedia system conveys information to users through different senses, responsible for displaying graphics, playing sounds, and providing haptic feedback to users.
Issues in the system can degrade user experiences in the multisensory perception from VR apps.

\subsubsection{\textbf{Implications}}
\label{subsubsec:rqs-multisensory-implications}

Considering VR users' great concern about multisensory perception, developers and researchers need to ensure the (a) interworking of different perception senses and (b) high-quality effects of every single sense.
According to our findings on multisensory perception, testing for VR apps' multimedia output is challenging, since it should simultaneously consider different levels of multisensory VR behavior (including graphics, audio, haptic feedback, etc.).

\textbf{Challenges.}
(a) \textit{Ensuring the interworking of diverse combinations of different perception senses} is challenging as it requires processing multisensory behavior and analyzing the correlations among combinations of different perceptions and application logic.
For instance, we can hardly detect unsynchronized effects of controller vibration, audio and graphics if we are not aware of the correlations among them.

(b) \textit{Ensuring high-quality effects of every single sense} is also challenging.
For ensuring high-quality graphics, existing work~\cite{paper:conf/ase/MacklonTVARPB22, paper:owl-eyes, paper:seenomaly} on traditional software already shows the difficulty of obtaining the expected displaying results (ground truth) and comparing them with actual results.
The problem can be even harder for VR applications as users have high expectations for graphic quality.
Ensuring high-quality audio and haptic effects is also not easy due to the difficulty of test oracle construction and verification: capturing and analyzing these two types of effects during dynamic testing is harder than graphics, and the two types of effects are also strongly correlated with application logic. 
Moreover, existing testing methods may not be applicable for application behavior in different modalities.

Few existing works on testing multimedia behavior can be directly adopted for quality assurance of multisensory perception, since most existing works only focus on ensuring high-quality graphics of traditional software~\cite{paper:glib-gui-oracles, paper:appflow-gui-testing2, paper:widget-gui-testing3}.
We provide several suggestions towards possible solutions as follows.

\textbf{Possible solutions.} 
(a) \textit{Metamorphic testing}.
One possible approach 
is the widely-used metamorphic testing~\cite{paper:metamorphic-testing}.
Researchers can design VR-specific metamorphic relations regarding the influencing factors we find (Figure~\ref{fig:factor-taxonomy}) to alleviate the test oracle problem.
(b) \textit{Multimodal deep learning (for ensuring the interworking of multiple senses).}
Graphics, audio, haptic effects and application code are in different modalities.
In order to detect low-quality practices of interworking, researchers can explore how to leverage multimodal deep learning methods to represent, align and fuse the output data of different senses in VR applications, and train a model to act as the test oracle. 
Multimodal deep learning~\cite{paper:multimodal-learning} has already shown its power for tasks involving multiple modalities, such as video retrieval~\cite{paper:multimodal-video-retrival}, emotion recognition~\cite{paper:multimodal-emotion-recog}, and decision making of autonomous systems~\cite{paper:multimodal-autodriving}.

\stepcounter{subimpcounter}
\begin{implicationbox}
	\begin{implication}
       Ensuring the quality of multisensory perception, i.e., ensuring (1) the interworking of diverse combinations of different perception senses and (2) the high-quality effects of every single sense, is challenging.
       This is due to the difficulties of (a) capturing/processing multisensory behavior, and (b) analyzing the correlations among combinations of different perceptions and application logic. 
       Developers and researchers may leverage metamorphic testing and multimodal learning approaches to address the challenges.
	\end{implication}
\end{implicationbox}

\subsection{User-Friendly Interaction Mechanism}
\label{subsec:rqs-interaction}

\subsubsection{\textbf{Detailed User Concerns}}
\label{subsubsec:rqs-interaction-user-concerns}
Interaction with VR applications is a lot different from traditional software, since users need to use special devices and their body movements to control their avatars and other objects in a 3D world.
Users favor the sense of presence and immersion brought by such revolutionized human-computer interaction.
However, they also suffer from issues regarding the usability, flexibility, and reliability of VR's new interaction mechanisms.
More than 14.1\% (219/\informativeReviewAmount) reviews in our labeled dataset are related to these issues.

\textbf{Usability of Interaction Mechanisms.}
For usability, users' complaints concentrate on three sub-categories. 
First, users are bothered when their VR applications are not easy to operate and interact with (poor \textbf{operability}).
Second, users also complain that they get confused when the interaction mechanisms cannot be easily learned or inferred, especially for beginners (poor \textbf{learnability}).
A user put down this comment: ``\textit{The interaction way is not intuitive and not explicitly stated,
	I spent my first five hours in the game swinging my arms around like an idiot before I discovered I 
	have to point and click}''.

Moreover, users complain when VR applications cannot be used normally by people with special needs, e.g., the mobility-limited, hearing-impaired, or vision-impaired group (poor \textbf{accessibility}).
Accessibility issues exist in traditional software as well.
However, such problems become a lot more serious and difficult to address in VR applications, which often rely on users' physical movement and multiple senses for interactions.
A user commented on one application of poor accessibility support for limited mobility: ``\textit{I must play from a sitting position, due to significant injury. The game makes me painfully reach either too high or too low to comfortably play}''.

\stepcounter{subfidcounter}
\begin{findingbox}
	\begin{finding}
       VR applications rely on users' physical movement and multiple senses for interactions, making accessibility problems more severe and harder to resolve as compared with traditional software.
	\end{finding}
\end{findingbox}

\textbf{Flexibility of Interaction Mechanisms}.
Users often complain when the interaction mechanisms 
are too restricted without flexibility.
For example, limited mobility may make users feel less free to interact with VR applications, thus decreasing their sense of control and immersion.
Here is a review containing such a complaint: ``\textit{lacking of free movement displeases me greatly, doesn't have any flexibility for people who prefer free movement or smooth turning}''.

\textbf{Reliability of Interaction Mechanisms}.
Users expect their interactions with VR applications to be stable in process and precise in results.
Their sense of realization and even comfort could be damaged otherwise, e.g., as complained in ``\textit{the interaction is imprecise and wonky to an immersion-breaking degree}''.

\stepcounter{subfidcounter}
\begin{findingbox}
	\begin{finding}
	    Users expect that VR applications should be easy to interact with, can be used normally by a wide range of people (even those with special needs), and the interaction mechanisms should be flexible, reliable, and easily learnable.
	\end{finding}
\end{findingbox}

\subsubsection{\textbf{Major Influencing Factors}}
\label{subsubsec:rqs-interaction-factors}

\textbf{Movement Mechanisms Design.}
\textit{Movement mechanisms} specify how users control their virtual avatar to take actions and move in virtual environments of VR applications.
There are more than 100 divergent movement mechanisms~\cite{paper:movement-mechanisms-MTI17, paper:movement-mechanisms-AHCI19, paper:movement-mechanisms-CHI21}, including room-scale natural locomotion (users' physical actions and movements within the physical space are tracked and replicated in virtual environments), teleportation (users use controller pointing and clicking to move instantly from one location to another within virtual environments), arm swing (user's arms move in a swinging motion to indicate movement in virtual environments like moving forward), etc.
Bad design choices or poor implementation of movement mechanisms can adversely affect the operability, learnability, accessibility, flexibility, and reliability of user experience on interaction mechanisms.
For example, as we have introduced above, teleportation is a widely implemented movement mechanism.
A user complained about the inflexibility of interaction mechanisms caused by the bad teleportation design of a VR shooting game: ``\textit{The main problem is the reliance upon teleportation. We can't move while shooting a gun. Why can't we just run by pushing the touchpad forward?}''.
Compared with traditional software, VR movement mechanisms
depend more on users' physical movements, such as height changes, turning around, etc.
How well and how much developers involve users' physical movements affect user experience on the quality attributes of interaction mechanisms, especially accessibility and flexibility.
More involvement of physical movements tends to (a) degrade the accessibility of interaction mechanisms for mobility-limited users, and (b) increase the flexibility of interaction mechanisms.

\textbf{Control Mechanisms Design.}
\textit{Control mechanisms} specify how users interact with and control all virtual objects other than their avatars.
Bad designs of control mechanisms degrade user experiences a lot on the quality attributes of interaction mechanisms.
For example, counterintuitive and too complicated designs induce poor operability and learnability.

\textbf{Real-Life Motion Tracking Implementation.}
As introduced in Section~\ref{subsec:vr-interaction}, VR applications track the position of head-mounted displays, handheld controllers, users' hands, etc., in the real world through position-tracking equipment.
Imprecise and unstable tracking implementation harms the reliability of interaction mechanisms.
Here's a complaint on imprecise tracking implementation
``\textit{BAD THINGS - Head/controller tracking sometimes doesn't track correctly, my arms fly 20 meters around me for the entire game}''.
Besides the three main influencing factors of interaction mechanisms, users also complain about factors like poor implementation of multimedia systems, poor support of configurations, and lacking of tutorials, which cause user-unfriendly interaction mechanisms.
For instance, poor implementation of multimedia systems can hurt the accessibility for sensory-impaired users.

\stepcounter{subfidcounter}
\begin{findingbox}
	\begin{finding}
       According to user reviews, the unsatisfactory interaction mechanisms are mainly caused by three factors: movement mechanisms design (for movement of user's virtual avatar), control mechanisms design (for interaction with other virtual objects), and real-life motion tracking implementation.

	\end{finding}
\end{findingbox}

\subsubsection{\textbf{Implications}}

Developers should pay more attention to developing and testing the interaction mechanisms, to ensure the relevant quality attributes that users care about most (
operability, learnability, accessibility, flexibility, and reliability).
As discussed in Section~\ref{subsubsec:rqs-interaction-user-concerns}, \textbf{accessibility} issues are more severe for VR applications than 
other software.
Let's take it as an example to elaborate on possible solutions for quality assurance.

\textbf{Quality assurance during development}. 
As introduced in Section~\ref{subsubsec:rqs-interaction-factors}, accessibility can be influenced badly by two factors, i.e., involvement of physical body movements and multisensory system implementation.
After analyzing the users' compliments about accessibility, we find that an effective method for improving accessibility is implementing sufficient options for users with impairments to configure to their special needs.
The options can be implemented from the perspectives of the two influencing factors:
(a) Developers are suggested to implement options for mobility-limited users by replacing inaccessible physical actions (e.g., users' height changes, turning around, etc.) with doable ones (e.g., pressing controller buttons).
For example, the popular VR game \textit{Half-Life: Alyx}~\cite{website:half-life-alyx} allows users to configure the \textit{height adjust options}, which users can configure to raise or lower their positions using controllers.
Users appreciate a lot on its design of accessibility options.
(b) It is hard to implement accessibility options for accessibility issues of severe impairment of one sense individually.
However, as introduced in Section~\ref{subsubsec:rqs-multisensory-userconcerns}, we find that in VR applications, different sense perceptions can interwork with each other to deliver information to users together.
It opens new possibilities for VR applications to mitigate such issues through apportioning information delivered via one sense to others.
For example, spatial audio and haptic effects can be provided to assist users' visual perception when they are about to hit onto virtual objects. 
Besides, developers should add accessibility options for other milder impairments, such as light sensitivity.

\begin{figure}[t!] 
	\centering 
	\includegraphics[width=\columnwidth]{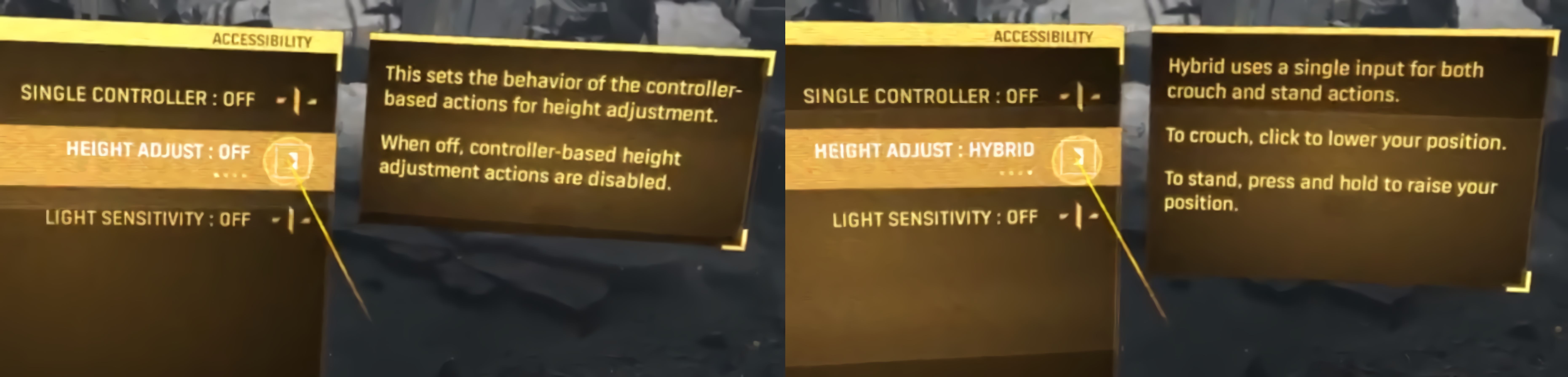} 
	\caption{Accessibility Options of Half-Life Alyx~\cite{website:half-life-alyx}, with \textit{height adjust} mode disabled on the left and \textit{hybrid height adjust} mode enabled on the right.}
	\label{fig:accessibility-options}
	       \vspace{-1.5em}
\end{figure}

\textbf{Quality assurance during manual testing}. 
Users complain a lot about insufficient accessibility testing, e.g., ``\textit{The game makes me painfully reaching either too high/low. Maybe there were few mobility-limited testers involved}''.
It would be better that the development team can anticipate and discuss all possible inaccessibility situations from the perspectives of each perception sense (including graphics, audio, haptic feedback, etc.) and each dimension of physical mobility for interaction mechanisms.
Then they can (a) hire a wider range of testers in different situations, or (b) mock the inaccessible scenarios by restricting testers' capability of perception senses or mobility, to verify the accessibility support during manual testing.
\stepcounter{subimpcounter}
\begin{implicationbox}
	\begin{implication}
       For improving accessibility, developers are suggested to implement sufficient accessibility options for users to configure, by (1) replacing inaccessible actions with doable ones for mobility-limited users, and (2) apportioning information delivered via one sense to others for sensory-impaired users, etc.
       During testing, developers are suggested to ensure accessibility support from the same two perspectives by (a) hiring a wider range of testers, or (b) mocking the inaccessible scenarios.
	\end{implication}
\end{implicationbox}
Test input generation for the movement mechanisms design and control mechanisms design is crucial for (a) ensuring operability, learnability, accessibility, flexibility and reliability of VR interaction mechanisms; and (b) exploring the application and ensuring other quality attributes that movement mechanisms and control mechanisms have influences on. 
However, the complicated and diverse movement mechanisms bring challenges to it, making it more difficult to generate effective test inputs as compared with traditional software.
We discuss the challenges and several possible solutions as follows.

\textbf{Huge input space (search space).} 
User reviews on the movement mechanisms and control mechanisms show that the input space for VR interaction is huge, 
as users report that diverse VR devices are used for interaction besides traditional devices like keyboards.
For example, the VR input space includes changing the position and orientation (both of which lie in the three-dimensional real number field) of head-mounted displays and two controllers, as well as controlling thumbsticks and pressing buttons on the controllers.
This makes the search space of prospective dynamic VR testing methods much more huge compared to traditional software. 
Existing testing approaches for traditional software with limited search space (typically keystrokes on keyboards and mouse clicking) can be ineffective in searching VR input consecutively in such a huge search space. 
Hence, researchers are suggested to propose VR-specific input searching methods for assuring quality attributes of interaction mechanisms.

\textbf{The effective search space differs under diverse movement mechanisms.}
As described in Section~\ref{subsubsec:rqs-interaction-factors}, we learn from reviews that effective user inputs for virtual movement under diverse movement mechanisms are significantly different.
A series of interaction sequences may make successful virtual movement under one type of movement mechanism (e.g., position movement of head-mounted displays and controllers can make expected virtual movement under room-scale natural locomotion), but fail to move under a different type of movement mechanisms (e.g., such position movement can hardly make any effect on movement under teleportation). 
Directly generating test inputs in the whole search space of all movement mechanisms seems to bypass this challenge. 
Considering the extremely large search space, it is challenging to perform prospective dynamic VR testing efficiently and effectively, which needs to explore as many functionalities as possible.

Several \textbf{possible solutions} are as follows:
\begin{itemize}[leftmargin=*, topsep=2pt, itemsep=2pt]
   \item \textbf{(a) Testing process isolation.}
   There exist several VR applications
   to change the movement mechanisms of other applications, e.g., \textit{Natural Locomotion}~\cite{website:natural-locomotion}.
   These applications inspire us that we can divide VR testing into two processes, and test the implementation of movement mechanisms (introduced above) and other functionalities separately.
   While testing other functions for a large number of VR applications, we can bypass the original implementation of movement mechanisms through a customized one (e.g., a customized SteamVR driver with general movement mechanisms for SteamVR applications). 
   This method is effective for large-scale testing.
   \item \textbf{(b) Subject categorization.}
   We can learn from user reviews that although there exist a large number of movement mechanisms, some of them rely on similar mechanics, which makes them share a common input space. 
   Hence, we can categorize the testing subjects of VR applications according to their movement mechanisms, and group those subjects with the same input space to test them together. 
   This idea is not effective enough for large-scale testing when lacking metadata for movement mechanisms, but is more convenient otherwise.
\end{itemize}

\stepcounter{subimpcounter}
\vspace{-0.9em}
\begin{implicationbox}
	\label{implication:interaction-researcher}
	\begin{implication}
       The complicated and diverse VR interaction mechanisms bring challenges to generating effective test inputs for the movement mechanisms and control mechanisms, making it hard to ensure quality attributes.
       The challenges can be addressed or bypassed by testing process isolation, subject categorization and VR-specific input searching methods.
	\end{implication}
\end{implicationbox}

\begin{figure}[t!] 
	\centering 
	\includegraphics[width=0.6\columnwidth]{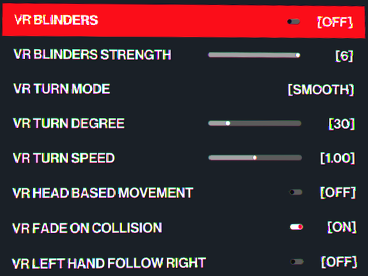} 
	\caption{Example of Comfort Options~\cite{website:hitman3}}
	\label{fig:comfort-options}
\end{figure}

\subsection{Immersivity}
\label{subsec:rqs-immersivity}

\subsubsection{\textbf{Detailed User Concerns}}

By nature, VR applications aim to provide more immersive experiences with 3D digital worlds.
This makes immersivity a unique quality attribute of VR experiences, which can hardly be achieved in traditional software. 
Thus, it is valued by VR users, discussed by around 12.7\% (197/\informativeReviewAmount) reviews in our labeled dataset. 
Immersivity consists of three sub quality attributes, including reasonability, submergence, and explorability:

\textbf{Reasonability}.
Users usually complain when the functionalities and logic of VR applications don't have a reasonable and coherent foundation (e.g., associated with real life or the general application logic).
Such unreasonable functions (e.g., strange physical phenomena/effects, imprecise positional tracking of movements) hurt users' sense of immersion.
Here is a complaint about imprecise tracking: ``\textit{The gun angle is tilted upwards, so your hands in VR and your hands in real life don't match up, killing any immersion I have}''.
\textbf{Submergence}.
Unlike 2D software, for VR applications, users are unsatisfied when they are still isolated behind a screen, cannot feel immerged/surrounded by virtual worlds and are detached from real life.
Here is a review that complains about poor submergence: 
``\textit{There are restrictions put in place that stop me from moving around freely in real life. This is a huge immersion breaker. You're telling me if I wanna step forward, I have to press buttons instead of just walking?}''.

\textbf{Explorability}.
Another major quality attribute of immersivity that users complain about is explorability.
Users expect to interact with objects and explore the virtual worlds in different combinations of routes and dimensions (e.g., both horizontally and vertically).
For example, here is a user's complaint on a VR application because of unsatisfactory explorability: ``\textit{this game has less explorability and is much more linear than its predecessors}''.
Poor design of object interaction and scene exploration can hurt users' immersion.

\stepcounter{subfidcounter}
\begin{findingbox}
	\begin{finding}
       Users expect VR applications to provide immersive experiences that can let them be detached from real life and submerge (submergence) in a coherent and reasonable virtual world (reasonability), where they can interact and explore in various dimensions (explorability).
	\end{finding}
\end{findingbox}

\subsubsection{\textbf{Major Influencing Factors}}
\label{subsubsec:rqs-immersivity-factors}

\textbf{(a) Movement Mechanisms Design - Involvement of Body Movement.}
Users complain when their body movements are not well involved in their interaction with VR applications.
For example, whether a sufficient amount of body parts are involved, how large the motion amplitude and movement area (movable range in the real world, possibly specified before using the app) are, how natural the body movements are integrated into interaction mechanisms, etc.
How well developers design the involvement of body movement affects users' immersive experiences.
For example, here is a complaint from users: ``\textit{(the app) uses only controllers and rarely involves any body movement, making you feel you're not really there}''.

\textbf{(b) Content Diversity.}
The mechanically generated and exactly same content ruin users' immersivity, giving them strong feelings that they are in digital worlds.
It is better to make VR applications more diverse and variable.
This will greatly increase users' sense of immersion, and can be achieved by procedural generation, handcrafting, etc.

\textbf{(c) Physics Design.}
\textit{Physics design} refers to the design of physics phenomena and effects in VR applications.
The physics design should at least create a reasonable and coherent virtual environment, preventing harms to users' sense of immersion.

\stepcounter{subfidcounter}
\begin{findingbox}
	\begin{finding}
       The design of movement mechanisms and control mechanisms have a huge influence on 
       immersivity.
       Multimedia system implementation and involvement of body movement affect reasonability and submergence. 
       Content diversity is an influencing factor for reasonability and explorability.
       Physics design is another influencing factor for reasonability.
	\end{finding}
\end{findingbox}

\subsubsection{\textbf{Implications}}
\label{subsubsec:immersivity-implications}

Ensuring the sub quality attributes of immersivity is nontrivial, since it is difficult to measure the degree of users' sense of reasonability, submergence and explorability.
We discuss several solutions to mitigate this problem:

\textbf{(a) Factor testing.} The problem can be converted into testing and improving the corresponding influencing factors we find (involvement of body movement, content diversity, physics design). 

\textbf{(b) Immersivity degree estimation.}
Designing questionnaires to let users rate their experiences~\cite{paper:questionnaire-game-chi15, paper:questionnaire-mobile-chi10} is a typical research method to quantify hard-to-measure attributes, e.g., the immersivity degree.
However, we find that most user reviews discussing immersivity only express general feelings with vague and subjective expressions, which means it may be hard for users themselves to realize the exact immersivity degree they experience, not to mention the prospective comparison between users and between applicaitons.
Many reviews mention that users experience different sentiments and feelings in different situations of immersivity, which makes it possible to leverage the electroencephalogram (EEG)~\cite{paper:eeg} to measure and record the electrical activity of users' brain and quantitatively estimate the degree of immersivity.

\stepcounter{subimpcounter}
\begin{implicationbox}
	\begin{implication}
       Researchers and developers can ensure immersivity through (1) testing the involvement of body movement, content diversity, physics design, etc., and (2) estimating the degree of immersivity leveraging EEG.
	\end{implication}
\end{implicationbox}

\begin{figure}[t!] 
	\centering 
		\includegraphics[width=0.6\columnwidth]{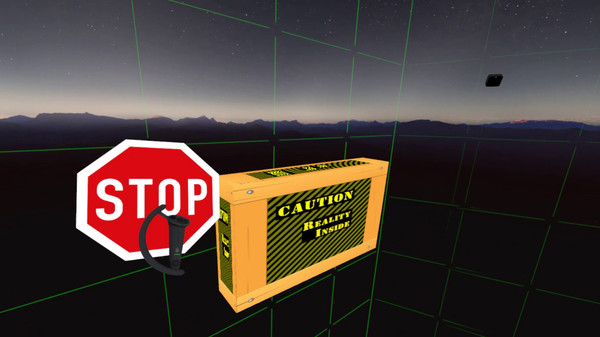} 
		\caption{Warnings for Real-Life Walls and Objects~\cite{website:stop-sign-vr-tools}}
		\label{fig:boundary-warnings}
\end{figure}

\subsection{Comfort \& Safety}
\label{subsec:rqs-comfort}

\subsubsection{\textbf{Detailed User Concerns}}

Comfort \& safety is an important and special quality attribute that traditional software developers and fresh VR developers rarely pay attention to.
Such issues are discussed in around 8.2\% (127/\informativeReviewAmount) reviews.

\textbf{Comfort.}
\textbf{(a) Neural Comfort.}
A common series of user complaints on comfort issues are related to neural feelings, such as motion sickness, nausea, and dizziness caused by neural stimuli like conflicts between perception and cognition in the brain.
One complaint example is: ``\textit{it is incredibly nauseating to have your head stop in the game but continue to move in the real world}''.

\textbf{(b) Physical Comfort.}
Users report another category of comfort issues caused by physical pains.
Typical physical pains result from inappropriate designs of interaction mechanisms.
Here is an example: ``\textit{holding down the two grip buttons was making my hands cramp}''.

\textbf{Clear Boundaries and Low Effects to Real Worlds (CBLE).}
While using VR applications, users are isolated in virtual worlds and tend to lose their real-world perception.
In such cases, it is important for VR applications (or the underlying platforms) to maintain clear boundaries between virtual worlds and nearby real environments (e.g., walls, objects, etc.), and lower the unintentional effects/damages in the real world under the actions users only intend to conduct in virtual worlds.
Here is a complaint about unclear boundaries: ``\textit{The only problem for me is that I can't see (the boundary of) my play area, and I accidentally bumped into real-life walls or objects}''.

\stepcounter{subfidcounter}
\begin{findingbox}
	\begin{finding}
       Lots of users experience and complain about neural discomfort, physical discomfort, and unclear boundaries to the real world in VR applications.
       Such issues bring severe consequences to users' physical health and safety.
	\end{finding}
\end{findingbox}

\subsubsection{\textbf{Major Influencing Factors}}
\label{subsubsec:rqs-comfort-factors}

\textbf{(a) For neural comfort.} When using VR applications, users' brains recognize the perception from the multimedia system during their virtual interactions in virtual worlds, while their body thinks they are just doing the actual real-life movement.
This causes conflicts between perception and cognition, thus resulting in neural comfort.
Recall that teleportation is a specific type of movement mechanism (Section~\ref{subsubsec:rqs-interaction-factors}).
A user stated his/her dissatisfaction with neural comfort due to poor movement mechanisms design that ``\textit{one thing wrong with this game is that teleportation is awful because it pulls you quickly in a sprint motion to the spot you are teleporting, which causes me lots of nausea}''.
Poor interaction stability or precision caused by the tracking implementation would make the situation worse. 
Note that low performance and poor functional suitability~\cite{standard:iso-25010-square-2011} also have knock-on effects on neural comfort.

\textbf{(b) For physical comfort and CBLE.} 
Users' complaints about the relevant issues mainly result from inappropriate designs of interaction mechanisms, especially real-life motion tracking implementation and movement mechanisms design (involvement of physical body movement).

\stepcounter{subfidcounter}
\begin{findingbox}
	\begin{finding}
       Users' complaints indicate that comfort issues are often caused by poor design or implementation of movement and control mechanisms, real-life motion tracking, multimedia system, configurations, and physics design as well as low performance.
	\end{finding}
\end{findingbox}

\subsubsection{\textbf{Implications}}
\label{subsubsec:comfort-implications}

We make a contrastive analysis of the user compliments and complaints about comfort \& safety issues, and find that such issues can be mitigated through the following perspectives.

\textbf{Ensuring comfort.} Developers are encouraged to: 

(a) Provide better implementation of the corresponding influencing factors, such as making less intense virtual movement,
involving more real-life body movement to lead the virtual interactions, more thorough designs of UI, etc.

(b) Provide comfort options for users to configure to their needs, such as the ways and attributes of virtual turning and movement, etc.
Figure~\ref{fig:comfort-options} shows an example in \textit{Hitman 3}~\cite{website:hitman3}.

(c) Make better assurance of the quality attributes which have knock-on effects on comfort, including performance efficiency~\cite{standard:iso-25010-square-2011}, multisensory perception, functional suitability~\cite{standard:iso-25010-square-2011} and so on.
For example, stable frames per second, lower latency, and fewer stuttering effects make users feel more comfortable.
Users complain that low quality of these quality attributes discomfort them. 
One example is that ``\textit{
	Performance-wise, the dropped frames on my GPU induce motion sickness over a long session}''.

(d) Conduct comfort testing besides functional testing before release. 

\textbf{Ensuring CBLE.} Developers are suggested to warn users when they are near the boundaries of the pre-specified play area and nearby obstacles (walls, objects, etc.).
One such example with many users' compliments is demonstrated in Figure~\ref{fig:boundary-warnings}.
The application shows {\color{applegreen}green grids} in the place of real-world walls and ceilings, and a {\color{alizarin}warning sign} will pop out if the user is about to bump into walls.
It also warns users about real-life objects with {\color{carrotorange}yellow boxes}.

\textbf{Pattern-based automated testing for comfort \& safety.}
Currently, the most 
popular VR testing approach is manual testing, which requires actual experiments involving human testers~\cite{website:vr-playtesting-guide}. %
Though effective, the cost of manual testing is high when testing a large number of VR applications with complicated logic.
The problem is more complicated for comfort \& safety testing as it may cause neurally or physically bad conditions to testers, and brings ethical issues.
This dilemma opens a research opportunity to design automated testing approaches for comfort \& safety issues.
During review analysis, we find that users tend to discuss the detailed causes to their discomfort or insecure experiences, e.g., the detailed experience of bad movement mechanisms, stuttering graphics, or unsynchronized motion tracking.
Based on that, researchers can collect detailed user reports about discomfort or insecure issues users have experienced, to learn bad patterns that hurt user experience on comfort \& safety.
Such knowledge can be used to construct tools to dynamically explore in VR applications and detect known bad patterns for comfort \& safety automatically.

\stepcounter{subimpcounter}
\begin{implicationbox}
	\begin{implication}
       Developers can mitigate the comfort \& safety issues through better implementation of the corresponding influencing factors, adding comfort options, 
       and implementing boundary warning modules for play area bounds and real-world objects.
       Due to the difficulty of manual testing, researchers are suggested to design automated testing approaches for comfort \& safety issues based on bad pattern collection and detection.
	\end{implication}
\end{implicationbox}

\textred{\subsection{Questionnaire-Based Supplementary Evidence}}

\textred{
To examine how practitioners and users assess the coverage and relevance of our VR software quality taxonomy, we distributed an anonymous questionnaire across multiple developer and user forums (Reddit, Discord, Meta Community, etc.). The survey solicited participants’ views on VR software quality attributes and their influencing factors, along with respondents’ development backgrounds. We ultimately received 43 valid responses. The English questionnaire can be accessed at: \href{https://forms.gle/82CFnrQw8pQMnk72A}{https://forms.gle/82CFnrQw8pQMnk72A}
}

\textred{
Among the 43 responses, 38 were in Chinese and 5 in English. The cohort comprises 34 participants with development experience, including 7 who have specifically worked on VR software and 16 with more than three years of professional practice. Their roles span development, testing, and project management. Respondents range in age from 10--19 to 50--59, with the largest group aged 20--29. This convenience sample adds perspectives beyond the review corpus, but its small size and concentration in Chinese-speaking communities limit generalization.
}  

\begin{figure}[t!] 
 	\centering 
 	\includegraphics[width=\columnwidth]{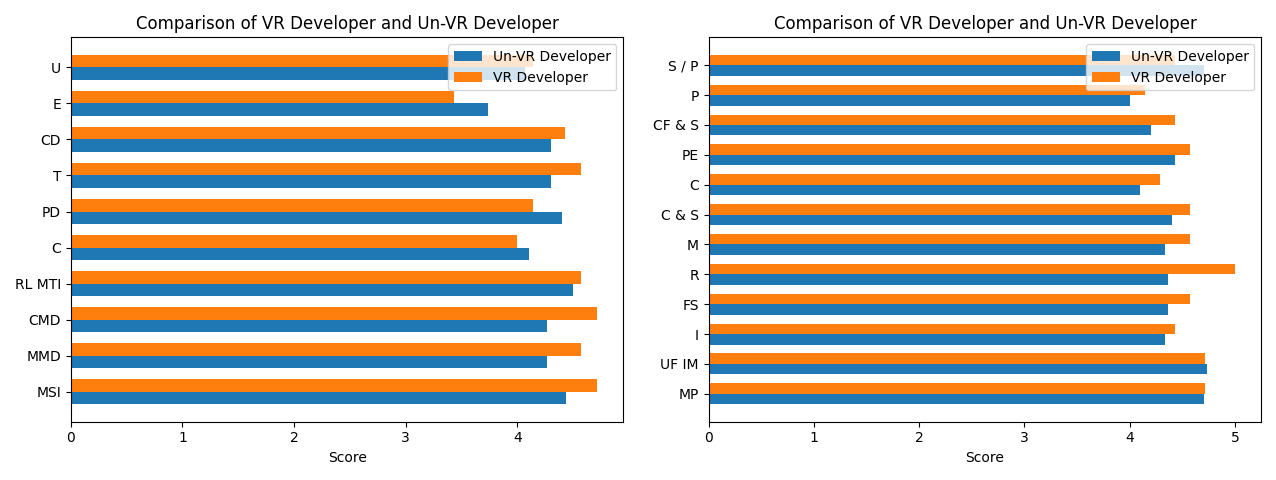} 
	\caption{\textred{Mean importance ratings by VR development experience}}
 	\label{fig:score}
 \end{figure}

\textred{
Figure \ref{fig:score} presents the mean importance ratings (on a 5-point scale) assigned by respondents with and without VR development experience to each quality attribute and influencing factor. Across most taxonomy items, both groups assigned high average ratings, generally near or above 4.0. Respondents with VR development experience rated several VR-specific attributes and factors higher, although this pattern is not uniform across all items. We therefore use the questionnaire as supplementary face- and content-validity evidence for the taxonomy, rather than as standalone proof of statistical robustness.
}

\textred{
In addition to the survey above, we included questions in our questionnaire to capture respondents’ subjective opinions and expectations regarding the future development of VR software technologies. For instance, we designed questions such as, “If you think that the rapid development of VR technology affects the different dimensions of the above VR software quality attributes, how do you think it will affect them?” and “If you think that the rapid development of VR technology affects the above different dimensions of factors affecting VR software quality attributes, what reasons do you base your considerations on?”
}

\textred{
The collected responses show that many respondents expect VR software quality concerns to evolve as the technology matures. Several respondents anticipated that hardware and software advances may reduce some current limitations, while also introducing new quality requirements as VR functionality expands and adoption becomes more widespread. For example, one respondent noted, “I believe that with the rapid development of technologies such as artificial intelligence, VR will inevitably integrate with these technologies in future application scenarios, leading to new standards and paradigms that will influence the requirements for new VR attributes.” Another respondent highlighted, “With the development of cross-platform solutions like OpenXR, portability is likely to become increasingly important. Additionally, since VR software can collect user privacy data in more dimensions compared to conventional software, privacy and security attributes may also gain greater significance.”
}

\textred{
Responses from participants without VR technology backgrounds also show that some expectations are framed at a perceptual or speculative level. For instance, one respondent suggested, “New technologies might mitigate issues such as security concerns,” while another remarked, “As the VR ecosystem becomes richer, the importance of third-party extensions will gradually increase; regarding motion tracking, if VR can be controlled directly through thoughts in the future, motion detection may become less important.” We interpret these comments cautiously: they do not establish technical feasibility, but they help characterize how non-specialist users reason about future VR quality attributes such as security and motion tracking.
}

\begin{figure*}[t!]
\centering
\includegraphics[width=\textwidth]{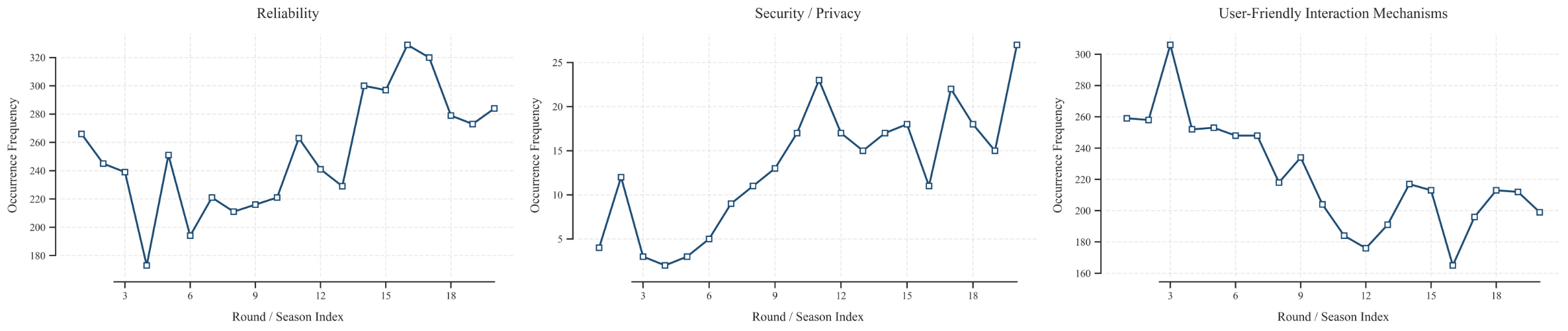}
\caption{\textred{Quarterly occurrence counts of selected quality labels in the temporally sampled reviews}}
\label{fig:temporal}
\end{figure*}

\textred{\subsection{Analysis of Temporal Effects}}
\textred{
The collected reviews span nearly a decade, from 2015 to 2025. Given the rapid evolution of VR hardware and software during this period, we examined whether user-perceived software quality concerns vary over time. Separately from the 6,925-sentence updated-corpus sample, we randomly drew 2,000 reviews for each of 20 quarters between June 2019 and June 2024; iOS App Store reviews were excluded because usable timestamps were unavailable. These temporal samples were annotated using the multi-LLM majority-voting pipeline followed by human review. We then tracked quarterly occurrence counts for selected labels, including User-Friendly Interaction Mechanisms, Reliability, and Security / Privacy, as illustrated in Figure~\ref{fig:temporal}. The figure provides descriptive temporal evidence; we do not claim statistical significance without a separately reported test and effect size.
}

\textred{
(a) User-Friendly Interaction Mechanisms: Mentions of User-Friendly Interaction Mechanisms show an overall downward pattern across the sampled quarters, with fluctuations in the middle and later periods. This pattern indicates that interaction-related concerns became less frequent in this sampled review stream, but the figure alone does not establish why the decline occurred.
}

\textred{
(b) Reliability: Reliability mentions fluctuate across the sampled quarters and increase in several later periods. The observed pattern is consistent with reliability remaining a persistent user concern, but it should not be interpreted as evidence that specific emerging technologies caused the change.\\
}
\textred{
(c) Security / Privacy: Security / Privacy has much lower absolute counts than the other two labels, but it appears more often in several later sampled quarters than in the earliest ones. This descriptive pattern suggests growing visibility of privacy- and security-related concerns in the sampled reviews, while the underlying causes require additional evidence beyond the figure.
}

\textred{
Taken together, these descriptive trends show that the relative attention paid to different quality labels is not static over time. In this sample, interaction-mechanism mentions decline overall, Reliability remains frequent and increases in later periods, and Security / Privacy remains rare but becomes more visible in some later quarters. These observations motivate temporal sensitivity in future VR review mining, while stronger claims about changing user priorities would require formal trend tests and qualitative follow-up.
}

\textred{\subsection{Accessibility-Related Review Subset}}
\textred{
To investigate VR software quality issues that are explicitly connected to accessibility and special-needs contexts, we first retrieved potentially relevant reviews by keyword matching (e.g., limited, disability, left-handed). After applying multi-LLM majority voting followed by human filtering, we retained 675 high-confidence samples. Because this procedure identifies review content rather than verified reviewer identities, we refer to the result as an accessibility-related subset instead of a confirmed population of special-needs users. Each sample was then re-labeled using the same multi-LLM + human pipeline, and the resulting subset was analyzed descriptively.}

\begin{figure}[t!] 
 	\centering 
 	\includegraphics[width=\columnwidth]{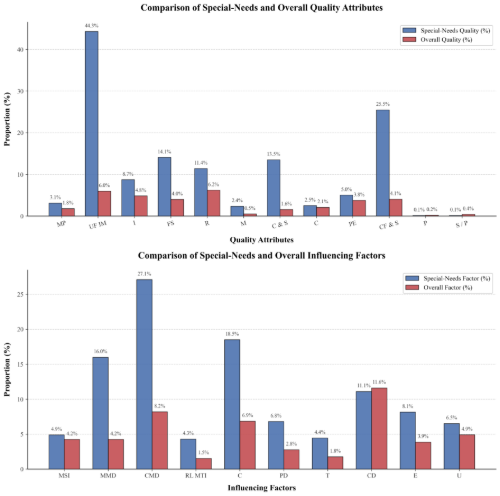} 
	\caption{\textred{Comparison of accessibility-related and overall review samples}}
	 \label{fig:accessibility-comparison}
 \end{figure}

\textred{
We compared the accessibility-related subset with the overall sample to examine whether the relative proportions of the taxonomy labels differ descriptively between the two groups. Figure~\ref{fig:accessibility-comparison} presents the results.}

\textred{
The accessibility-related subset has visibly higher proportions for User-Friendly Interaction Mechanisms, Comfort \& Safety, Functional Suitability, and Performance Efficiency than the overall sample. For influencing factors, Control Mode, Content, Motion/Movement Design, and several interaction-related factors are also more prominent in this subset. These differences should be interpreted as descriptive contrasts; statistical significance would require a separate proportion test and effect-size report.
}

\textred{(a) User-Friendly Interaction Mechanisms and Configuration Flexibility \& Simplicity: }

\textred{
Many reviews in the accessibility-related subset describe difficulty with complex interaction patterns or spatial requirements. These comments suggest that some users need interaction schemes that are simple, configurable, and compatible with seated play or limited movement. Evidence for this preference appears in reviews such as the following:}

\textred{
“Suitable for wheelchair users; size calculated automatically. Horizontal and vertical walls can be disabled, as can tunnels … three difficulty levels: beginner, advanced, and expert … Please always consider wheelchair users. Just sit in a chair and test what is possible. And think of older adults in retirement homes who want to use an Oculus: why should I stop at 80?”
}

\textred{
(b) Immersivity: Some accessibility-related reviews discuss immersive experiences as beneficial in contexts where users report physical or psychological constraints. Such comments suggest that a sense of presence can be important for this subset, but they should not be generalized as clinical evidence. As a reviewer stated, “This app has been genuinely beneficial. Diagnosed with PTSD and generalized anxiety disorder, I often struggle to initiate meditation or even regulate my breathing. Even when conventional techniques fail, the application transports me to a serene, aesthetically compelling virtual space where I can simply sit and inhabit an alternate world.”
}

\textred{
(c) Comfort \& Safety: Reviews in this subset also show that interaction or configuration assumptions can create physical strain for users who cannot comfortably perform the required posture or movement. One reviewer, who experiences knee pain, noted that the title’s cover system mandates a “knees-on-floor posture for most of the session,” making the experience “uncomfortable or even prohibitive” for individuals with lower-limb impairments.
}

\textred{
(d) Tutorial: Some reviews in the accessibility-related subset also point to the need for clearer onboarding guidance. As one older reviewer succinctly remarked, “Can someone give an old guy help on how to start the game?”
}

\textred{
Based on the preceding descriptive analysis, we propose the following evidence-informed guidelines for VR developers:
}

\textred{
Developers should support seated and limited-movement use, configurable reach and boundary settings, alternatives for information conveyed through only one sensory channel, and clear onboarding for unfamiliar interaction schemes. These requirements should be evaluated with people who use the relevant accessibility features; the review subset identifies recurring needs but does not determine which implementation is appropriate for every application.
}

\begin{figure*}[t!]
\centering
\includegraphics[width=\textwidth]{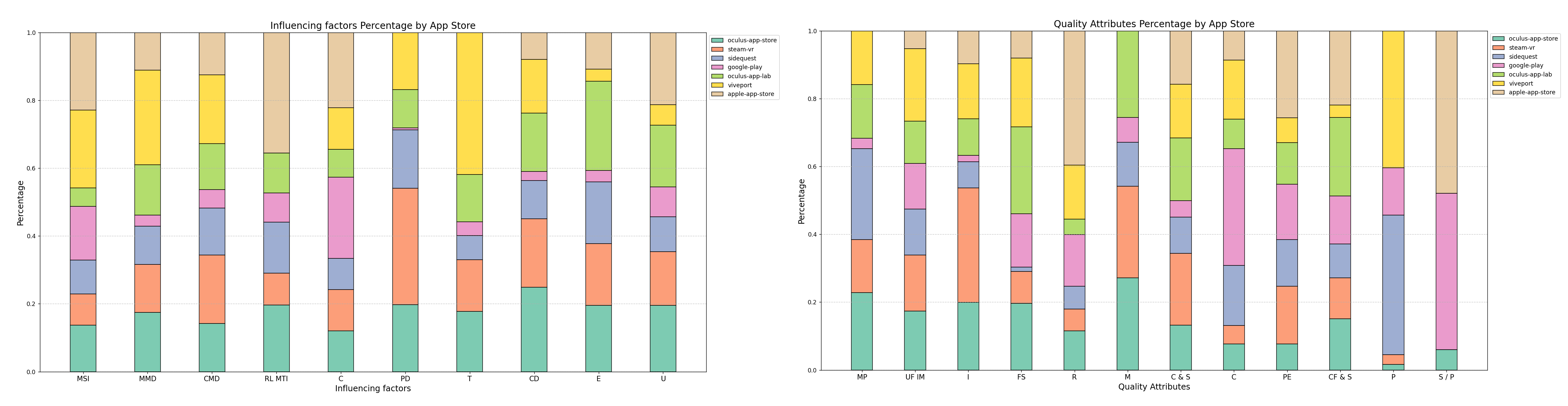}
\caption{\textblue{Platform composition of each quality attribute and influencing factor}}
\label{fig:platform}
\end{figure*}

\textblue{\subsection{Cross-Platform Quality Perceptions}}
\textblue{
Because VR apps on different storefronts exhibit distinct characteristics and serve varied user bases, we analyzed how each quality attribute and influencing factor is distributed across platforms. Figure~\ref{fig:platform} shows, for each taxonomy label, the relative contribution of reviews from each platform; each stacked bar is normalized within a label and sums to 1. This visualization therefore supports claims about platform composition within each label, not platform-normalized prevalence rates. The distributions vary across labels, with Security / Privacy and Compatibility containing relatively larger shares of mobile-store reviews than many other labels.
}

\textblue{
(a) Security / Privacy: In Figure~\ref{fig:platform}, Security / Privacy is composed largely of mobile-store reviews, especially Google Play, with smaller contributions from several other platforms. This composition indicates where the Security / Privacy labels in our sample came from; it does not by itself establish per-platform prevalence. Representative remarks include:}
\textblue{
• Data exfiltration: “Well, I know for a fact that Meta uses and sends my data to third parties. Which is a big yikes.”\\
• Malware: “Phone picked up virus in app; should not be in Play Store.”\\
• Over-privileging: “Asks about a lot of permissions, and even if you uninstall the app it gets re-installed automatically.”\\
• Regulatory non-compliance: “This application is not compliant with European privacy regulations.”\\
• Unauthorized billing: “It seems to be a scam … charged subscription billed to my phone bill without permission.”}

\textblue{
The mobile-store composition is consistent with the broader privacy and security concerns raised in the representative reviews, including data sharing, malware, excessive permissions, regulatory compliance, and unauthorized billing. However, the figure does not identify the technical causes of these concerns. Explanations involving mobile operating systems, third-party SDKs, advertising networks, or update practices should therefore be treated as hypotheses for follow-up analysis rather than conclusions from this dataset alone.
}

\textblue{
(b) Compatibility: Compatibility labels also contain a comparatively large share of mobile-store reviews in Figure~\ref{fig:platform}. The representative reviews illustrate several recurring forms of compatibility concern:}
\textblue{
• App simply fails to launch: “Well I can’t get this thing to work with Galaxy Note 8 … it won’t even bring up the correct app screen, then it won’t connect to the steering wheel.”\\
• Resolution or UI mis-match: “Only 1080p resolution can play this app. I tried 2880×1440 but the picture was totally blank, while the audio played … please add opacity for the UI or at least make it wider.”\\
• Cross-device fragmentation: “Game isn’t bad, but broke it within the first minute … also needs to be compatible with other VR apps, as most phones aren’t compatible with Google’s Cardboard app.”}

\textblue{
These comments are consistent with well-known mobile VR challenges such as device heterogeneity, screen-resolution differences, sensor variation, and differences among headset/controller profiles. Still, the current figure and review examples do not isolate root causes, so we frame this interpretation as plausible context rather than causal evidence.
}

\textblue{
Overall, Figure~\ref{fig:platform} shows that the platform composition of quality-related reviews varies by taxonomy label. Mobile storefronts contribute a larger share to Security / Privacy and Compatibility labels than to many other labels, while dedicated VR storefronts contribute more heavily to other concerns. A platform-normalized prevalence analysis would be needed before claiming that a given platform has a higher rate of a label among its own reviews.
}

\begin{figure*}[t!]
\centering
\includegraphics[width=\textwidth]{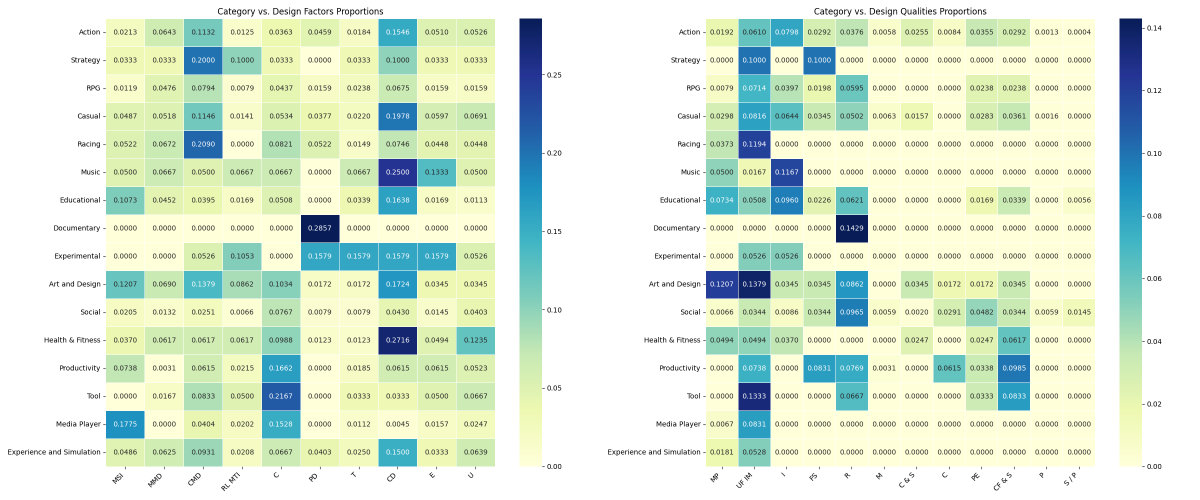}
\caption{\textblue{Genre-normalized proportions of quality attributes and influencing factors}}
\label{fig:genres}
\end{figure*}

\textblue{\subsection{Genre-Based Taxonomy of Quality Perceptions}}
\textblue{
The app-level metadata retrieved from VR storefronts includes genre labels, which provide an additional lens for examining quality perceptions. Because each marketplace uses its own taxonomy, we normalized the labels into 16 genres and grouped them into consumer-oriented, professional-oriented, and ambiguous/hybrid proxies, as shown in Table~\ref{tab:vr-categories}. These proxies describe store metadata; they do not verify consumer or enterprise deployment. For each genre, we calculated the proportion of reviews mentioning each quality attribute and influencing factor and visualized the results as a heat map in Figure~\ref{fig:genres}. The heat map is descriptive: it identifies genre-associated concentrations in the sample but does not establish that genre membership causes a quality concern.
}

\begin{table}[htbp]
    \centering
    \caption{VR Application Categories}
    \label{tab:vr-categories}
    \resizebox{0.45\textwidth}{!}{%
    \begin{tabular}{p{3cm} p{6cm}}
        \toprule
        \textbf{Category} & \textbf{Applications} \\
        \midrule
        \textblue{Consumer-oriented} & Action, Strategy, RPG, Casual, Racing, Music, Social, Media Player \\
        \cdashlinelr{1-2}
        \textblue{Professional-oriented} & Educational, Health \& Fitness, Productivity, Tool \\
        \cdashlinelr{1-2}
        Ambiguous/Hybrid & Documentary, Experimental, Art and Design, Experience and Simulation \\
        \bottomrule
    \end{tabular}%
    }
\end{table}

\textblue{
Multisensory Perception: Art \& Design has one of the highest proportions of Multisensory Perception labels in Figure~\ref{fig:genres}. This pattern is consistent with reviews that discuss coordinated visual, auditory, and experiential qualities in creative VR applications. One review, for example, praises how the graphics, music, and sound effects work together to create a relaxing experience.}

\textblue{
Comfort \& Safety and Configuration Flexibility \& Simplicity: Health \& Fitness shows relatively high proportions for Configuration Flexibility \& Simplicity and movement- or control-related concerns, while Functional Suitability is not prominent for this genre in the heat map. This pattern is consistent with the physical and motion-intensive nature of exercise-oriented VR applications, where boundaries, posture, tracking, and configuration can affect the user experience. As a reviewer remarked, "It does its job, allows you to have a workout. Only annoying things are that it constantly puts stuff outside of boundaries which results in sometimes hitting a wall or window or having to reach out of the boundary. The sprinting detection thing isn’t the best but it does the job. Otherwise amazing app and helps to keep fit."
}

\textblue{
Security / Privacy: Social applications show one of the higher Security / Privacy proportions in Figure~\ref{fig:genres}, although the absolute values for this label are small across genres. This pattern is consistent with the networked and identity-bearing nature of social VR applications, but the heat map alone does not establish higher actual privacy risk. One reviewer stated, “Don't give the app your info; they can access your wifi network and pupil dilations as well.”
}

\begin{figure*}[t!]
\centering
\includegraphics[width=\textwidth]{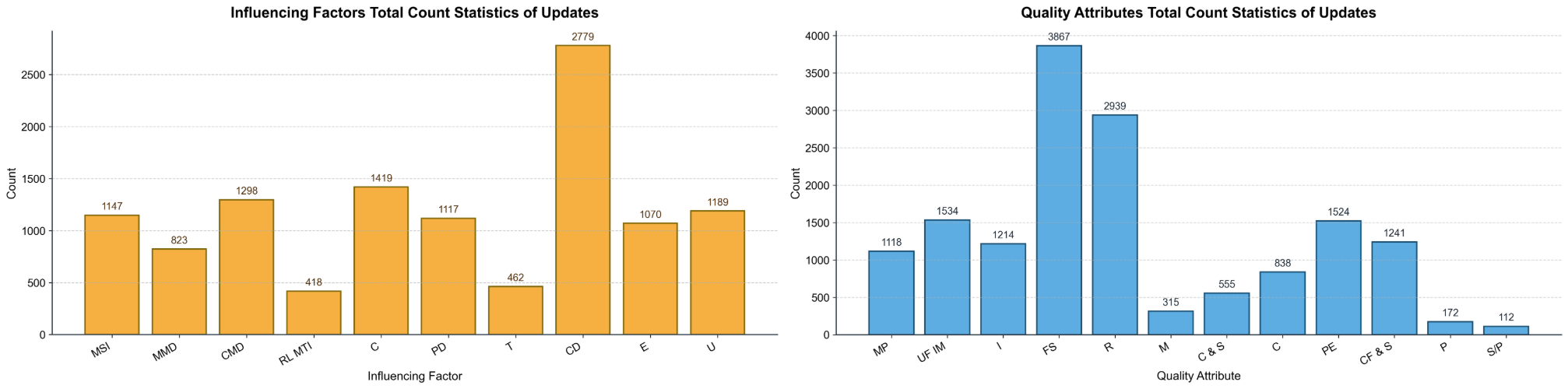}
\caption{\textblue{Label counts in update logs}}
\label{fig:update}
\end{figure*}

\textblue{\subsection{Quality Concerns in Developer Update Logs}}
\textblue{
To study VR software quality issues from a developer-facing perspective, we collected update logs from four app stores: Meta Horizon Store, SideQuest, Steam, and iOS App Store. Update logs were not available on VIVEPORT or Google Play. We collected 49,319 update logs from 5,084 applications. To facilitate analysis, we selected the top 500 applications by number of user reviews and analyzed their 11,951 update logs. An LLM assigned Quality Attribute and Influencing Factor labels to these logs, and Figure~\ref{fig:update} presents the resulting counts. Because these labels were not produced by the original manual focused-coding procedure, classification error is an additional limitation. Moreover, the figure reports raw counts, so comparisons with user reviews require normalized denominators and are interpreted descriptively.}

\textblue{
In the update logs, Content Diversity is the most frequent influencing factor label (2,779 instances), followed by Content (1,419), Control Mode (1,298), Usefulness (1,189), Multisensory Information (1,147), Performance Deficiency (1,117), and Environment (1,070). This distribution indicates that update logs often document additions or changes to content, features, assets, and related application material. For instance, updates such as "Today we are releasing Chapter 1. Entanglement, which includes 9 tracks in both MP3 (320K) and FLAC formats" illustrate the type of content-expansion activity captured by this label. These counts show what developers record in release histories; they do not by themselves prove developer priorities, user retention effects, or direct mitigation of user complaints.}

\textblue{
For quality attributes, Functional Suitability (3,867 instances) and Reliability (2,939 instances) are the two most frequent update-log labels. Other common labels include User-Friendly Interaction Mechanisms (1,534), Performance Efficiency (1,524), Configuration Flexibility \& Simplicity (1,241), and Immersivity (1,214). Examples from the logs, such as "Fixed recent problem preventing creating or updating new Workshop items" or "Updated auto-saving to be more reliable," illustrate how release histories frequently record functionality and stability changes. These findings support a descriptive conclusion that update logs emphasize feature/function changes and reliability fixes more than labels such as Portability or Security / Privacy.}

\textblue{
Taken together, the update-log analysis offers a complementary developer-facing view of VR software maintenance. Release histories frequently mention content additions, functionality changes, and reliability fixes, while some concerns that are prominent in user reviews may require separate review-mining or issue-tracking evidence to connect them to concrete updates. A stronger claim about whether developers fixed issues raised in reviews would require linking individual reviews to subsequent version-history entries for the same application and release window.
}

\section{Threats to Validity}
\label{sec:threats}

\subsection{Internal Validity}
\textblue{The procedures of taxonomy construction and manual labeling present conjoined threats to internal validity. Automated aspect extraction methods can miss review expressions, merge distinct concerns, or surface noisy phrases, and preprocessing plus manual relevance filtering cannot establish whether a review is genuine, independent, or representative of an actual end-user experience. The resulting taxonomy should therefore be interpreted as a structured model of quality concerns observable in our public review corpus, not as an exhaustive inventory of all VR software quality concerns.}

\textblue{We mitigated these threats by using multiple aspect extraction techniques as inputs for qualitative analysis, conducting multi-round open coding~\cite{book:open-coding-16} with five evaluators, and then applying focused coding~\cite{book:sage-qda-focused-coding} to refine and label sampled review sentences. The taxonomy evolved during these iterations, so conventional inter-rater reliability statistics over a fixed codebook are not available and would not faithfully describe the early open-coding process. Instead, our mitigation relies on collaborative review of every evolving taxonomy, independent first-pass labeling during focused coding, repeated discussion of discrepancies, explicit refinement of category definitions and boundaries, reanalysis after taxonomy updates, and consensus after \manualAnalysisIterationNum focused-coding iterations. Figure~\ref{fig:quality-taxonomy} and Figure~\ref{fig:factor-taxonomy} report the residual \textit{other} labels, but these counts indicate only that the final codebook covered most informative sampled sentences under our coding procedure; they do not prove category completeness.}

\textblue{The two analysis stages also involved partly different personnel: five evaluators constructed the taxonomy, while three experienced analyzers conducted focused coding, with one overlapping analyzer explaining the taxonomy definitions. This design reduced direct dependence on a single group, but it should be viewed as a consensus-based qualitative validation rather than an independent statistical validation. For transparency, the public artifact releases the review datasets, labeled dataset, high-level preprocessing and labeling workflow, analyzer guidelines, and taxonomy iteration materials. It does not include all intermediate aspect-extraction outputs, per-annotator labels, disagreement records, adjudication notes, or supplementary analysis data; consequently, those materials should not be claimed as publicly available replication artifacts.}

\subsection{External Validity}
\textblue{Our dataset collection process poses threats to external validity. The reviewers who post in public app stores are a proxy for VR app users, not the full population of VR users. Public reviewers may be unusually motivated, dissatisfied, experienced, or platform-specific, and we cannot guarantee their demographic background, device ownership, usage context, or relationship to the app developers. The English-only collection further concentrates the evidence in communities where English public reviews are available, so the taxonomy may underrepresent concerns expressed in other languages, markets, and cultural contexts.}

\textblue{The scope of our findings is also bounded by consumer-facing and publicly reviewable VR applications in the studied marketplaces. The results may not transfer directly to closed enterprise deployments, laboratory prototypes, specialized medical or training systems, unreleased applications, or future devices whose interaction modalities, sensing capabilities, safety constraints, and distribution channels differ from current public stores. Our platform-level coverage improves breadth within this public-marketplace setting, but it does not remove the consumer-market and time-period boundaries of the dataset.}

\textblue{A further limitation is that public app-store review systems do not provide a guaranteed reviewer identity or a reliable basis for detecting coordinated, incentivized, or otherwise manipulated reviews. Promotional astroturfing, brigading, duplicated campaigns, or malicious downvoting could affect observed frequencies. Our preprocessing and manual relevance filtering reduce obvious noise and exclude sentences that are not informative for the taxonomy, but they cannot verify reviewer genuineness, independence, or motivation. Therefore, the reported category proportions should be read as descriptive signals in the collected corpus rather than precise prevalence estimates for the underlying VR user population.}

\textblue{Finally, the supplementary questionnaire and metadata analyses should be interpreted as descriptive triangulation rather than causal evidence or general validation of the taxonomy. The questionnaire is a small convenience sample with linguistic and cultural concentration, and the metadata-based analyses are limited by fields available from public marketplaces. LLM-assisted labels in the updated-review and update-log analyses introduce additional model and prompt sensitivity; majority agreement plus human review reduces this risk for the updated-review sample, whereas the update-log counts retain the limitations of their LLM classifier. These analyses can show whether selected findings are consistent with another source of evidence, but they do not establish causal relationships between app characteristics, developer actions, and review concerns or guarantee generalizability beyond the sampled respondents, platforms, genres, collection period, or available metadata.}

\section{Conclusion}

\textred{VR quality assurance must account for the embodied, immersive, and multisensory consequences of software behavior. We therefore studied the quality concerns that reviewers express about deployed VR applications and the design and implementation factors associated with those concerns.}
\textred{Across \totalReviewAmount public reviews of \totalAppAmount applications, our semiautomatic analysis produced a hierarchical model of \qualityFirstLayerCategoryAmount quality attributes and \factorFirstLayerCategoryAmount influencing factors. The model captures both conventional attributes with VR-specific consequences and concerns rooted in multisensory perception, interaction, immersivity, and comfort and safety. Its links to movement, control, tracking, multimedia, physics, configuration, and content provide concrete targets for requirements, testing, and maintenance. Our supplementary analyses further show descriptive variation across time, marketplaces, genres, accessibility-related feedback, and update logs; these contextual results should not be interpreted causally. Together, the model and evidence offer a common vocabulary for cumulative research on VR software quality and a practical basis for prioritizing quality-assurance work.}
\textred{To facilitate inspection and follow-up research, we release the available datasets, taxonomy materials, and analysis results at \href{https://sites.google.com/view/vr-software-quality}{\color{blue}{https://sites.google.com/view/vr-software-quality}}.}

\clearpage

\balance
\bibliographystyle{IEEEtran}
\bibliography{xr-review}

\end{document}